\documentclass[10pt,twocolumn,twoside]{IEEEtran}

\usepackage{lipsum}
\usepackage{graphicx}
\usepackage{algorithm}
\usepackage[noend]{algpseudocode}
\usepackage{cite}
\usepackage{pgfplots}
\usepackage{bm}
\usepackage[nohyperlinks, nolist]{acronym}
\usepackage{amsfonts}
\usepackage{tikz-3dplot}
\usetikzlibrary{3d}

\definecolor{c1}{HTML}{f2ae72}
\definecolor{c2}{HTML}{f2e394}
\definecolor{c3}{HTML}{588c7e}
\definecolor{c4}{HTML}{d96459}
\definecolor{c5}{HTML}{e0e0e0}
\definecolor{c6}{HTML}{AA336A}
\definecolor{c7}{HTML}{C3B1E1}
\definecolor{c8}{HTML}{FFD580}
\definecolor{c9}{HTML}{FFF9A6}
\definecolor{c10}{HTML}{702963}
\definecolor{c11}{HTML}{000C66}
\definecolor{c12}{HTML}{D36F85}

\makeatletter
\newcommand{\vast}{\bBigg@{3.3}}
\newcommand{\Vast}{\bBigg@{5}}
\makeatother

\newcommand{\figfont}{\normalfont\fontsize{7}{7}\selectfont}

\makeatletter
\long\def\@makecaption#1#2{\ifx\@captype\@IEEEtablestring%
\footnotesize\begin{center}{\normalfont\footnotesize #1}\\
{\normalfont\footnotesize\scshape #2}\end{center}%
\@IEEEtablecaptionsepspace
\else
\@IEEEfigurecaptionsepspace
\setbox\@tempboxa\hbox{\normalfont\footnotesize {#1.}~~ #2}%
\ifdim \wd\@tempboxa >\hsize%
\setbox\@tempboxa\hbox{\normalfont\footnotesize {#1.}~~ }%
\parbox[t]{\hsize}{\normalfont\footnotesize \noindent\unhbox\@tempboxa#2}%
\else
\hbox to\hsize{\normalfont\footnotesize\hfil\box\@tempboxa\hfil}\fi\fi}
\makeatother

\ifCLASSOPTIONcompsoc
    \usepackage[caption=false, font=normalsize, labelfont=sf, textfont=sf, subrefformat=parens, labelformat=parens]{subfig}
\else
\usepackage[caption=false, font=footnotesize, subrefformat=parens, labelformat=parens]{subfig}
\fi

\newcommand{\code}[1]{{\fontfamily{qcr}\selectfont #1}}

\makeatletter
\renewcommand{\ALG@name}{\small Algorithm}
\makeatother
\makeatletter
\def\BState{\State\hskip-\ALG@thistlm}
\makeatother

\ifCLASSINFOpdf
\else
\fi

\usepackage{graphicx,url}
\usepackage[nosumlimits]{amsmath}
\usepackage{tabularx}
\usepackage{amssymb}
\usepackage{float}
\usepackage[colorlinks=true,linkcolor=black,citecolor=black]{hyperref}
\usepackage[capitalise]{cleveref}
\usepackage[nohyperlinks, nolist]{acronym}
\usepackage{multirow}
\usepackage{tikz}
\usepackage[none]{hyphenat}
\usepackage[flushleft]{threeparttable}
\usepackage[percent]{overpic}
\usetikzlibrary{positioning,chains}
\usetikzlibrary{shapes,arrows}
\usetikzlibrary{shapes.multipart}
\usepackage{siunitx}
\usepackage{tikz-timing}
\usepackage{pgfplots}

\usepackage{colortbl}
\usepackage{hhline}
\usetikzlibrary{svg.path}
\usepackage{scalerel}
\usepackage{mathrsfs}
\usepackage{transparent}
\usepackage{eso-pic}
\makeatletter
\newcommand*{\org@overidelabel}{}
\let\org@overridelabel\@verridelabel
\@ifpackagelater{acronym}{2015/03/21}{% v1.41
  \renewcommand*{\@verridelabel}[1]{%
    \@bsphack
    \protected@write\@auxout{}{\string\AC@undonewlabel{#1@cref}}%
    \org@overridelabel{#1}%
    \@esphack
  }%
}{% older versions
  \renewcommand*{\@verridelabel}[1]{%
    \@bsphack
    \protected@write\@auxout{}{\string\undonewlabel{#1@cref}}%
    \org@overridelabel{#1}%
    \@esphack
  }%
}
\makeatother

\usetikzlibrary{patterns}

\newcommand{\crefsub}[1]{Fig.~\subref*{#1}}
\crefformat{equation}{(#2#1#3)}

\crefmultiformat{equation}{(#2#1#3)}%
{ and~(#2#1#3)}{, (#2#1#3)}{ and~(#2#1#3)}
\crefrangeformat{equation}{(#3#1#4) to~(#5#2#6)}

\hypersetup{urlcolor=black}

\definecolor{orcidlogocol}{HTML}{A6CE39}
\tikzset{
    orcidlogo/.pic={
        \fill[orcidlogocol] svg{M256,128c0,70.7-57.3,128-128,128C57.3,256,0,198.7,0,128C0,57.3,57.3,0,128,0C198.7,0,256,57.3,256,128z};
        \fill[white] svg{M86.3,186.2H70.9V79.1h15.4v48.4V186.2z}
        svg{M108.9,79.1h41.6c39.6,0,57,28.3,57,53.6c0,27.5-21.5,53.6-56.8,53.6h-41.8V79.1z M124.3,172.4h24.5c34.9,0,42.9-26.5,42.9-39.7c0-21.5-13.7-39.7-43.7-39.7h-23.7V172.4z}
        svg{M88.7,56.8c0,5.5-4.5,10.1-10.1,10.1c-5.6,0-10.1-4.6-10.1-10.1c0-5.6,4.5-10.1,10.1-10.1C84.2,46.7,88.7,51.3,88.7,56.8z};
    }
}

\newcommand\orcidicon[1]{\href{https://orcid.org/#1}{\mbox{\scalerel*{
                \begin{tikzpicture}[yscale=-1,transform shape]
                \pic{orcidlogo};
                \end{tikzpicture}
            }{|}}}}

\pgfplotsset{compat=1.14} 

\begin{document}
\begin{acronym}[ANOVA]
    \acro{2D}{two-dimensional}
    \acro{3D}{three-dimensional}
    \acro{ACF}{auto-correlation function}
    \acro{AIC}{Akaike information criterion}
    \acro{AIR}{acoustic impulse response}
    \acro{ALS}{alternating least squares}
    \acro{APEVD}{approximate polynomial eigenvalue decomposition}
    \acro{AR}{augmented reality}
    \acro{ARI}{Acoustic Research Institute} 
    \acro{ASR}{automatic speech recognition}
    \acro{AuxIVA}{auxiliary function-based independent vector analysis}
    \acro{BIC}{Bayesian information criterion}
    \acro{BLSTM}{bidirectional long short term memory network}
    \acro{BNRQ}{block-based normalised Rayleigh quotient}
    \acro{BSD}{Bark spectral distortion}
    \acro{BSI}{blind system identification}
    \acro{BSS}{blind source separation}
    \acro{BURQ}{block-based unconstrained Rayleigh quotient}
    \acro{CCF}{cross-correlation function}
    \acro{CNN}{convolutional neural network}
    \acro{COLSUB}{coloured noise subspace-based speech enhancement}
    \acro{CPD}{canonical polyadic decomposition}
    \acro{CRNN}{convolutional recurrent neural network}
    \acro{CSD}{cross-spectral density}
    \acro{DBN}{deep belief network} 
    \acro{DFT}{discrete Fourier transform}
    \acro{DNN}{deep neural network}
    \acro{DRR}{direct-to-reverberant ratio}
    \acro{DTF}{directional transfer function}
    \acro{DoA}{direction of arrival}
    \acro{E-HMM}{evolutive hidden Markov model}
    \acro{ECG}{electrocardiography}
    \acro{EVD}{eigenvalue decomposition}
    \acro{F-norm}{Frobenius-norm}
    \acro{F0}[$F_{0}$]{fundamental frequency}
    \acro{FAR}{false alarm rate}
    \acro{FA}{false alarm}
    \acro{FCMCLMS}{frequency-domain constrained multichannel least mean squares}
    \acro{FFT}{fast Fourier transform}
    \acro{FFV}{fundamental frequency variation}
    \acro{FIR}{finite impulse response}
    \acro{FLMS}{fast least mean squares}
    \acro{FNMCLMS}{frequency-domain normalised multichannel least mean squares}
    \acro{FUMCLMS}{frequency-domain unconstrained multichannel least mean squares}
    \acro{FiNS}{filtered noise shaping}
    \acro{FwSegSNR}{frequency-weighted segmental signal-to-noise ratio}
    \acro{GAN}{generative adversarial network}
    \acro{GCC}{generalized cross-correlation}
    \acro{GCD}{greatest common divisor}
    \acro{GEVD}{generalised eigenvalue decomposition}
    \acro{GNN}{graph neural network}
    \acro{GSC}{generalised sidelobe canceller}
    \acro{GPU}{graphics processing unit}
    \acro{GWPE}{generalised weighted prediction error}
    \acro{HMM}{hidden markov model}
    \acro{HOS}{higher order statistics}
    \acro{HRTF}{head-related transfer function}
    \acro{HR}{HIT rate}
    \acro{ICA}{independent component analysis}
    \acro{IDFT}{inverse discrete Fourier transform}
    \acro{IFB}{independent frequency bin}
    \acro{IFFT}{inverse fast Fourier transform}
    \acro{IHM}{individual headset microphone}
    \acro{IIR}{infinite impulse response}
    \acro{ILRMA}{independent low-rank matrix analysis}
    \acro{IR}{impulse response}
    \acro{ISCLP}{integrated sidelobe canceller linear prediction Kalman filter}
    \acro{ISHT}{inverse spherical Harmonic transform}
    \acro{ITD}{interaural time difference}
    \acro{ILD}{interaural level difference}
    \acro{IVA}{independent vector analysis}
    \acro{KLT}{Karhunen-Lo\`eve transform}
    \acro{LMS}{least mean squares}
    \acro{LPC}{linear predictive coding}
    % \acro{LSD}{log spectral distance}
    \acro{LR}{learning rate}
    \acro{LReLU}[Leaky ReLU]{Leaky rectified linear unit}
    \acro{LSD}{log-spectral distortion}
    \acro{LSP}{line spectral pair}
    \acro{LS}{least squares}
    \acro{LTI}{linear time-invariant}
    \acro{M-SMD}{modified sequential matrix diagonalization}
    \acro{MAP}{maximum a posteriori}
    \acro{MCLMS}{multichannel least mean squares}
    \acro{MCN}{multichannel Newton}
    \acro{MCSUB}{multi-channel subspace}
    \acro{MCULMS}{multichannel unconstrained least mean squares}
    \acro{MDL}{minimum description length}
    \acro{MDP}{Markov decision process}
    \acro{MFCC}{mel-frequency cepstral coefficient}
    \acro{MHT}{multiple hypothesis tracking}
    \acro{MH}{multi-hit}
    \acro{MIMO}{multiple input multiple output}
    \acro{MINT}{multi-channel inversion theorem}
    \acro{MISO}{multiple input single output}
    \acro{ML}{machine learning}
    \acro{MLSVD}{multi-linear singular value decomposition}
    \acro{MMSE}{minimum mean square error}
    \acro{MNMF}{multi-channel non-negative matrix factorization}
    \acro{MS-SBR2}{multiple shift second-order sequential best rotation}
    \acro{MS-SMD}{multiple Shift sequential matrix diagonalization}
    \acro{MSE}{mean square error}
    \acro{MSE}{mean squared error}
    \acro{MSE}{mean squared error}
    \acro{MUSHRA}{multiple stimuli with hidden reference and anchor}
    \acro{MUSIC}{multiple signal classification}
    \acro{MVDR}{minimum variance distortionless response}
    \acro{MWC}{maximum weighted clique}
    \acro{MWF}{multi-channel Wiener filter}
    \acro{MWIS}{maximum weighted independent set}
    \acro{MaxDir}{maximum directivity index}
    \acro{NMF}{non-negative matrix factorisation}
    \acro{NPM}{normalised projection misalignment}
    \acro{NSRR}{normalized signal-to-reverberant ratio}
    \acro{NSV}{negative-side variance}
    \acro{OMWF}{oracle multi-channel Wiener filter}
    \acro{PCA}{principal component analysis}
    \acro{PEFAC}{pitch estimation filter with amplitude compression}
    \acro{PESQ}{perceptual evaluation of speech quality}
    \acro{PEVD}{polynomial eigenvalue decomposition}
    \acro{PHAT}{phase-transform}
    \acro{PMD}{polynomial matrix decomposition}
    \acro{PMUSIC}{polynomial multiple signal classification}
    \acro{PRTF}{polynomial relative transfer function}
    \acro{PSD}{power spectral density}
    \acro{RGB}{red, green and blue}
    \acro{RIR}{room impulse response}
    \acro{RL}{reinforcement learning}
    \acro{RLR}{regularised linear regression}
    \acro{RMS}{root mean square}
    \acro{RTF}{relative transfer function}
    \acro{SAR}{source-to-artefacts ratio}
    \acro{SBR2-T}{second-order sequential best rotation with tridiagonal reduction}
    \acro{SBR2}{second-order sequential best rotation}
    \acro{SBR2}{sequential best rotation algorithm}
    \acro{SCA}{sparse component analysis}
    \acro{SD}{standard deviation}
    \acro{SDM}{single distant microphone}
    \acro{SDN}{scattering delay network}
    \acro{SDR}{source-to-distortion ratio}
    \acro{SHT}{spherical harmonic transform}
    \acro{SH}{spherical harmonic}
    \acro{SIMO}{single input multiple output}
    \acro{SIR}{source-to-interference ratio}
    \acro{SISO}{single input single output}
    \acro{SLAM}{simultaneous localization and mapping}
    \acro{SMA}{spherical microphone array}
    \acro{SMD}{sequential matrix diagonalisation}
    \acro{SMD}{sequential matrix diagonalisation}
    \acro{SNR}{signal-to-noise ratio}
    \acro{SNR}{signal-to-noise ratio}
    \acro{SOS}{second order statistics}
    \acro{SP-MUSIC}{spatial polynomial MUSIC}
    \acro{SRGAN}{super-resolution generative adversarial network}
    \acro{SR}{super-resolution}
    \acro{SSP-MUSIC}{spatio-spectral polynomial MUSIC}
    \acro{STFT}{short-time Fourier transform}
    \acro{STOI}{short-time objective intelligibility}
    \acro{SURQ}{sample-based unconstrained Rayleigh quotient}
    \acro{SVD}{singular value decomposition}
    \acro{SegSNR}{segmental signal-to-noise ratio}
    \acro{T60}{reverberation time}
    \acro{T60}{reverberation time}
    \acro{TDE}{time-delay estimation}
    \acro{TDoA}{time difference of arrival}
    \acro{TKD}{Tucker decomposition}
    \acro{TT}{tensor train}
    \acro{ToA}{time of arrival}
    \acro{VAD}{voice activity detection}
    \acro{VAD}{voice activity detection}
    \acro{VR}{virtual reality}
    \acro{WPD}{weighted power minimum distortionless response}
    \acro{WPE}{weighted prediction error}
    \acro{i.i.d.}{independent identically distributed}
    \acro{log-MMSE}{log-minimum mean square error}
    \acro{HRIR}{head-related impulse response}
    % \acro{SD}{log-spectral distortion}
    \acro{PReLU}{parametric rectified linear unit}
    \acro{ReLU}{rectified linear unit}
\end{acronym}
\makeatletter
\title{HRTF upsampling with a generative adversarial\\ network using a gnomonic equiangular projection}\let\Title\@title

\author{
Aidan~O.~T.~Hogg\,\textsuperscript{\orcidicon{0000-0001-5501-7799}}, %~\IEEEmembership{Member,~IEEE,}
        Mads~Jenkins,
        He~Liu,
        Isaac Squires\,\textsuperscript{\orcidicon{0000-0003-1919-061X}},
        Samuel J. Cooper\,\textsuperscript{\orcidicon{0000-0003-4055-6903}},
        and~Lorenzo~Picinali\,\textsuperscript{\orcidicon{0000-0001-9297-2613}}
\thanks{
This study was made possible by support from SONICOM (www.sonicom.eu), a project that has received funding from the European Union’s Horizon 2020 research and innovation program under grant agreement No. 101017743.

A. O. T. Hogg is with both the Centre for Digital Music, Queen Mary University of London, 327 Mile End Road, London, E1 4NS, UK and the Audio Experience Design Group, Dyson Sch. of Design Engineering, Imperial College London, Exhibition Road, SW7 2AZ, UK (e-mail: aidan@aidanhogg.uk).

Lorenzo~Picinali is with the Audio Experience Design Group, Dyson Sch. of Design Engineering, Imperial College London, Exhibition Road, SW7 2AZ, UK (e-mail: l.picinali@imperial.ac.uk).

Mads Jenkins and He Liu were with the Audio Experience Design Group, Dyson Sch. of Design Engineering, Imperial College London, Exhibition Road, SW7 2AZ, UK. 

Isaac Squires and Samuel J. Cooper are with the Tools for Learning, Design and Research Group, Dyson Sch. of Design Engineering, Imperial College London, Exhibition Road, SW7 2AZ, UK (e-mail: i.squires20@imperial.ac.uk and samuel.cooper@imperial.ac.uk).

}% <-this % stops a space
% \thanks{Manuscript received April 19, 2005; revised August 26, 2015.}
}\let\Author\@author
\makeatother
% The paper headers
\markboth{IEEE/ACM TRANSACTIONS ON AUDIO, SPEECH, AND LANGUAGE PROCESSING}%
{A. HOGG \MakeLowercase{\textit{et al.}}: {\let \\ \empty {\Title}}}

\maketitle
% \IEEEpubid{\begin{minipage}{\textwidth}\ \\\\\\\\[12pt] \centering
%   This work has been submitted to the IEEE for possible publication. \\Copyright may be transferred without notice, after which this version may no longer be accessible.
% \end{minipage}} 
\begin{abstract}
An individualised \ac{HRTF} is very important for creating realistic \ac{VR} and \ac{AR} environments. However, acoustically measuring high-quality \acp{HRTF} requires expensive equipment and an acoustic lab setting. To overcome these limitations and to make this measurement more efficient \ac{HRTF} upsampling has been exploited in the past where a high-resolution \ac{HRTF} is created from a low-resolution one. This paper demonstrates how \acp{GAN} can be applied to \ac{HRTF} upsampling. We propose a novel approach that transforms the \ac{HRTF} data for direct use with a convolutional \ac{SRGAN}. This new approach is benchmarked against three baselines: barycentric upsampling, \ac{SH} upsampling and an \ac{HRTF} selection approach. Experimental results show that the proposed method outperforms all three baselines in terms of \ac{LSD} and localisation performance using perceptual models when the input \ac{HRTF} is sparse (less than 20 measured positions).
\end{abstract}


% \IEEEpeerreviewmaketitle
\acresetall
\sisetup{tight-spacing=true}

\section{Introduction}
\label{sec:intro} 

\IEEEPARstart{R}{emote} interaction has grown in use in recent years, however, there are still many unsolved problems with remote connectivity. A common issue is the lack of immersive audio in these virtual interactions. Immersive audio is what people experience in their everyday lives; some sounds are close, some are far away, some are moving, some are static, and all come from different directions. The loss of the acoustic spatial dimension, as well as the physical interactions with sound, can lead to the frustration people often feel when communicating remotely (e.g. \cite{Lokki2004}). This is immediately apparent in online meetings when multiple participants try to speak simultaneously or struggle to hear what others are saying. This problem worsens when some participants are present in person and others online. However, this need not be the case, and realistic immersive audio attempts to contribute to solving this problem by making the transition from real to virtual audio seamless.

Furthermore, improvements to online communication are not the only aim. In recent years, the prevalence of \ac{VR} and \ac{AR} devices, as well as \ac{3D} video games, auditory displays and hearing assistive devices \cite{Vickers2021}, has led to a demand for more realistic \ac{3D} audio rendering. Therefore, the need for better immersive audio solutions is becoming increasingly relevant to the modern world.

One way to achieve high realism in immersive audio is to place the listener in the centre of a large spherical loudspeaker array and play different sounds from different directions in space. Unfortunately, although this solution works well in an acoustic lab and could be used in cinemas and other large venues, it is not practical beyond these controlled settings. It is also costly and only works well for the small number of participants at the centre of the array. A more practical solution exploits the fact that humans have two auditory sensors (i.e. two ears); in theory, we should only need two speakers (i.e. in-ear headphones) to generate the correct sound at those sensors, performing what is commonly referred to as binaural (i.e. involving the two ears) spatialisation. 

\subsection{The matter of individualisation}

The main challenge with binaural spatialisation is understanding how sounds at the entrance of the two ear canals can be realistically generated to mimic real-world 3D audio accurately \cite{Wightman1989} and, more specifically, how this can be adapted for individual listeners. This individualisation has resulted in a large amount of research focusing on \acp{HRTF}, which capture the filtering effects related to the anatomy of different listeners. This filtering is caused by the sound wave being reflected and scattered off the head, torso, and pinnae before it enters the ear canal of a given listener. \acp{HRTF} are, therefore, able to capture interaural (i.e. the difference heard between the listener's two ears) and monaural localisation cues \cite{Blauert1983}. 
It is customary to refer to \ac{HRTF} when considering the \ac{IR} in the frequency domain and \ac{HRIR} in the time domain. In this paper, we also use the term \acp{HRTF} to refer to the complete set of \ac{IR} measurements corresponding to a full set of source positions around each listener.

It has been shown in the past that using non-individualised \acp{HRTF} can significantly affect the individual's sound source localisation accuracy~\cite{Stitt2019, Wenzel1993, Moller1996} as spectral cues are highly dependent on the listener's anatomy, particularly the shape of their pinnae \cite{Kahana2006}. This non-individualisation can also affect perceptual attributes such as externalisation, immersion, colouration, realism and relief/depth~\cite{Simon2016, Werner2016, Engel2019}. Furthermore, the choice of \ac{HRTF} can significantly impact an individual's ability to understand speech in a cocktail party scenario~\cite{Cuevas-Rodriguez2021}.

As a result, capturing individual listeners' personalised \ac{HRTF} remains an important area of active research. It has been shown in the past that many approaches can be deployed for this task of \ac{HRTF} individualisation, including acoustic measurements \cite{Engel2023}, \ac{3D} scans \cite{Ziegelwanger2015a},  modelling the morphological geometric information of the listener's ears \cite{Katz2001, Stitt2021} and selecting the best-fitting \ac{HRTF} from a database of already-measured ones. This best-fitting \ac{HRTF} selection is often made using morphology-based methods \cite{Geronazzo2019, Zotkin2003} or perceptual-based methods (e.g. using individual preference \cite{Katz2012} and/or localisation accuracy \cite{Kim2020, Zagala2020}). An overview of some of the most common methods can be found in \cite{Picinali2022}.

The acoustic measurement \cite{Carpentier2014} is still considered the gold standard of these different approaches. The downside to performing this acoustic measurement is the expensive custom setup required and the time it takes. This is because numerous \acp{IR} need to be measured around the individual, with numbers ranging anywhere from 200 to 3000 \cite{Engel2023}. This process can be sped up by taking advantage of interlaced sine sweeps \cite{Farina2000}, but this often only makes the elevation measurements faster. Other methods do exist \cite{Zotkin2006, Richter2016} that aim to improve the time performance of the \ac{HRTF} measurement, but the equipment specifications and cost are usually very high.

\subsection{Spatial upsampling of HRTFs}

To reduce the time required and the complexity of the \ac{HRTF} setup and to make the method scalable, spatial upsampling methods have been proposed in the past that can generate high-resolution \acp{HRTF}, i.e. \acp{HRTF} that contain many (normally over 300) \acp{IR} from many directions, from low-resolution \acp{HRTF}, i.e. \acp{HRTF} that include very few \acp{IR} from very few directions \cite{Zhong2014}. This process is commonly referred to as \ac{HRTF} upsampling and can be achieved using various approaches.

The most common \ac{HRTF} upsampling method is barycentric interpolation \cite{Hartung1999, Poirier-Quinot2018a, Cuevas-Rodriguez2019}. This method has been shown to produce good results when the \acp{HRTF} contain a relatively large number of \acp{IR} \cite{Gamper2013}, for example, with an angular distance of 10-15\textdegree between measurements; however, it becomes much less reliable when interpolating sparser measurements (e.g. each 30-40\textdegree). Another common approach is \ac{SH} interpolation \cite{Evans1998, Porschmann2019, Arend2021, Engel2022, Arend2023b}, but again results in poor reconstruction when the low-resolution \ac{HRTF} input is spatially sparse. This is because these methods rely on averaging between existing data points based on prior information. For example, barycentric interpolation uses the three nearest neighbours around the point to be interpolated to calculate the weighted average. Therefore, as the distances between the neighbours grow larger, the upsampling becomes more and more inaccurate.

More recently, \ac{ML} methods have started to become the focus of research on \ac{HRTF} personalisation. In the past, \ac{ML} techniques have been shown to be effective at estimating \acp{HRTF} from just the anthropometric measurements of the listener. In \cite{Chen2019}, a \ac{DNN}-based approach is used to synthesise a personalised \ac{HRTF} using the anthropometric features of the user and was able to achieve a performance of 3.2~dB \ac{LSD}. This approach consisted of using the encoder part of an autoencoder to reduce the dimensionality of the raw \acp{HRTF}, which can be used as a set of input features for training. This aims to minimise overfitting as \ac{HRTF} datasets are usually small. The decoder part of the autoencoder then estimates the \ac{HRTF} magnitudes using the output of a \ac{DNN} that is trained to output the latent representation given the anthropometric features and the target azimuth. This type of approach is explored further in \cite{Yao2022}, where two autoencoders are exploited: one that reduces the dimensionality of a feature vector containing the azimuth and anthropometric parameters and another that reduces the dimensionality of the magnitudes of the full \ac{HRTF} measurement. These two autoencoders are combined to estimate the \ac{HRTF} magnitudes from anthropometric features and achieve a performance of 4.3~dB \ac{LSD}.

It has also been shown in the past that autoencoders can be used to upsample low-resolution \acp{HRTF}. In \cite{Ito2022}, a method is proposed that uses an autoencoder and is an extension of a \ac{RLR} approach that makes use of the spherical wavefunction presented in \cite{Duraiswami2004}. This method's key feature is that it decomposes \acp{HRTF} into source position-dependent and source position-independent factors, i.e. the spherical wavefunction expansion and expansion coefficients, respectively. The autoencoder is conditioned on source positions and obtains the source-position-independent representation by using an aggregation module between the encoder and decoder, aggregating latent variables of a given source position. This approach was able to notably achieve 4.4~dB in \ac{LSD} when upsampling from 9 to 440 positions. Other \ac{ML} methods also exist that are able to perform the task of \ac{HRTF} upsampling, including \cite{Kestler2019}, which exploits a \ac{DBN}. This method accomplishes an \ac{LSD} average of less than 3~dB; however, results are only given for upsampling from 125 source positions to 1250, which is still relatively dense. 
Another method in \cite{Jiang2023} uses a \ac{CNN} and has been shown to yield a good performance of 4.4~dB \ac{LSD} when upsampling from 23 positions to 1250 and 3.8~dB when upsampling from 105 positions. However, in this case, the sphere is sliced into planes to create a \ac{2D} representation rather than considering the sphere as a whole. \ac{ML} techniques have also been used in the past as part of a postprocessing step of the \ac{SHT} interpolation \cite{Tsui2020}. 

\subsection{HRTFs and GANs - our proposed solution}

The main advantage of \ac{ML} approaches over traditional upsampling is that they can extrapolate patterns from the data rather than these patterns being hard-coded, making it possible to recreate the missing information in the sparse measurements using the knowledge learnt from a training set that contains many high-resolution \acp{HRTF}. The aim of this paper is to investigate the use of \acp{GAN} to tackle the \ac{HRTF} upsampling problem, explicitly looking at very sparse \ac{HRTF} measurements and offering insight into the practicality of this approach. In the past, \acp{GAN} have been successfully applied to many audio applications, including WaveGAN \cite{Donahue2018}, which applies \acp{GAN} to the unsupervised synthesis of raw-waveform audio. \acp{GAN} were also used in \cite{Eskimez2019} for speech super-resolution, which aims to upsample a given speech signal by generating the missing high-frequency content. These applications of \acp{GAN} are motivated by the fact that \acp{GAN} have been successfully applied to the task of upsampling photos \cite{Dong2016, Ledig2017} and astronomical images \cite{Schawinski2017} and are often referred to as \acp{SRGAN}. \acp{SRGAN} \cite{Goodfellow2014} are a family of \ac{ML} models characterised by the use of two networks that compete in an adversarial manner. These models have been shown to work well on upsampling very low-resolution images; however, apart from a pilot study \cite{Siripornpitak2022}, they have not been exploited for the task of \ac{HRTF} upsampling.

A novel approach is proposed here using the \ac{SRGAN} framework, as introduced by \cite{Ledig2017}, to allow the generation of accurate high-quality \acp{HRTF} from sparsely measured ones, thus making this personal acoustic data available faster and at a lower cost, albeit requiring a small number of measurements anyway. The paper builds on a pilot study that was undertaken in \cite{Siripornpitak2022}, which explored using an \ac{SRGAN} for upsampling \acp{HRTF} across single planes in space, e.g. the horizontal, median and vertical planes. This limitation is overcome in the study presented here, where the full \ac{3D} \ac{HRTF} is employed for the SRGAN training and prediction. The next steps to further validate this technique, extend it, and ultimately integrate it within a tool to be openly released are outlined at the end of the paper.

The first challenge that was tackled was to transform the original \ac{HRTF} data into something more suitable for the \ac{SRGAN}, and this was achieved through various transformations and resampling operations. The main transformation is that of a gnomonic equiangular projection \cite{Ronchi1996, Rancic1996, Purser1998}, often referred to as a cubed sphere. The reason this type of projection was selected is that it does not produce singularities at the poles \cite{Purser2011}, and the distortion is quasi-uniform over the whole sphere \cite{Putman2007}.

The transformed \ac{HRTF} was then used to train the \ac{SRGAN}, for which an updated loss function was actually designed. Finally, an evaluation was carried out by spatially upsampling a certain number of low-resolution \acp{HRTF} and comparing the results with various benchmark techniques (e.g. barycentric and spherical harmonics interpolations). The comparison relied on both signal-level metrics and model-based perceptual evaluations.  

This paper is structured as follows: 
% \cref{sec:backgroud} gives a background overview on \acp{HRTF}. 
\cref{sec:method} introduces the method, including the pre- and post-processing steps along with the \ac{GAN} architecture. \cref{sec:experimental_setup} explains the experimental setup, that is, the data used, how the \ac{GAN} was trained, and an explanation of the baselines that were used for comparison. In \cref{sec:results}, spectral and perceptual model-based results are presented. Finally, \cref{sec:conclusion} provides the conclusions drawn.

% \section{HRTF Backgroud}
% \label{sec:backgroud}

% \subsection{Head-related transfer functions (HRTFs)}
% \section{Review of Polynomial MUSIC}
% \label{sec:fundamental_frequency_investigation} \noindent

\section{Method} 
\label{sec:method} 

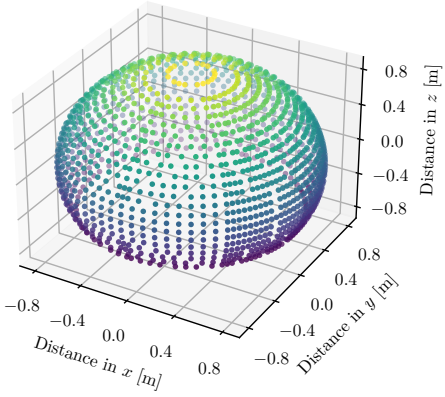
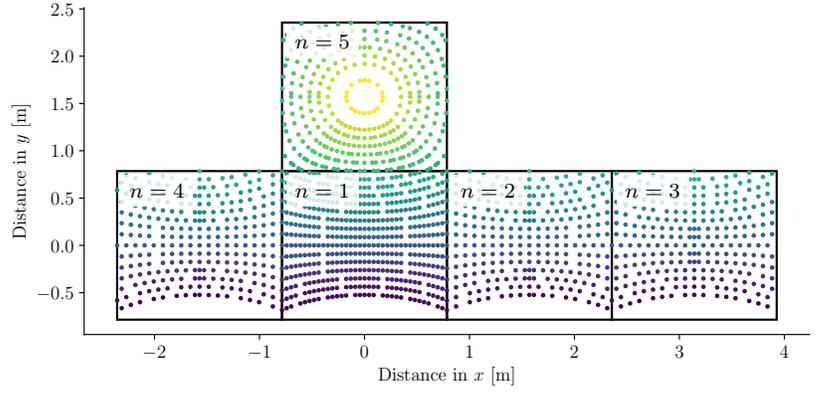
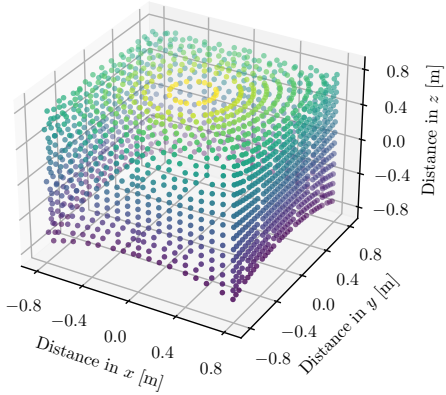
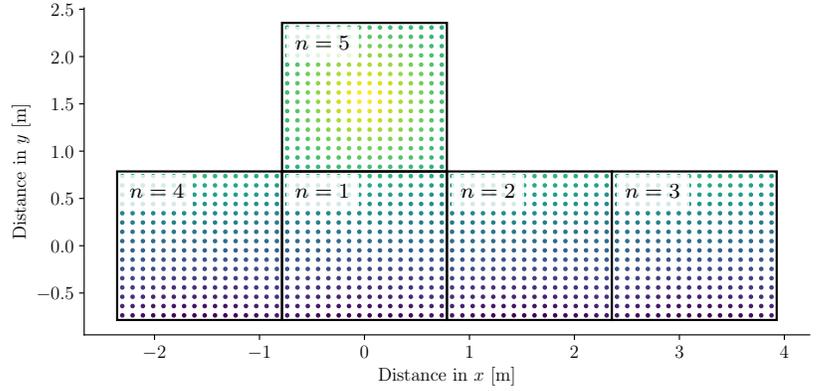
\begin{figure*}[tb]
\centering
  \setcounter{subfigure}{0}
   \hfill
  \subfloat[Original positions of the \acp{IR}.\label{fig:orginal_data_sphere}]{%
  \includegraphics[width=0.35\linewidth]{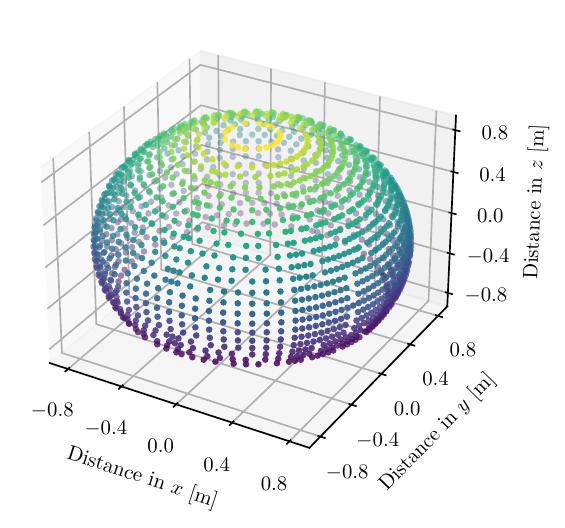}
}
  \hfill
  \setcounter{subfigure}{2}
  \subfloat[Flattened cube of projected positions of the \acp{IR}\label{fig:projected_data_flattened}]{
  \begin{tikzpicture}
   \node (img) {\includegraphics[width=0.6\linewidth]{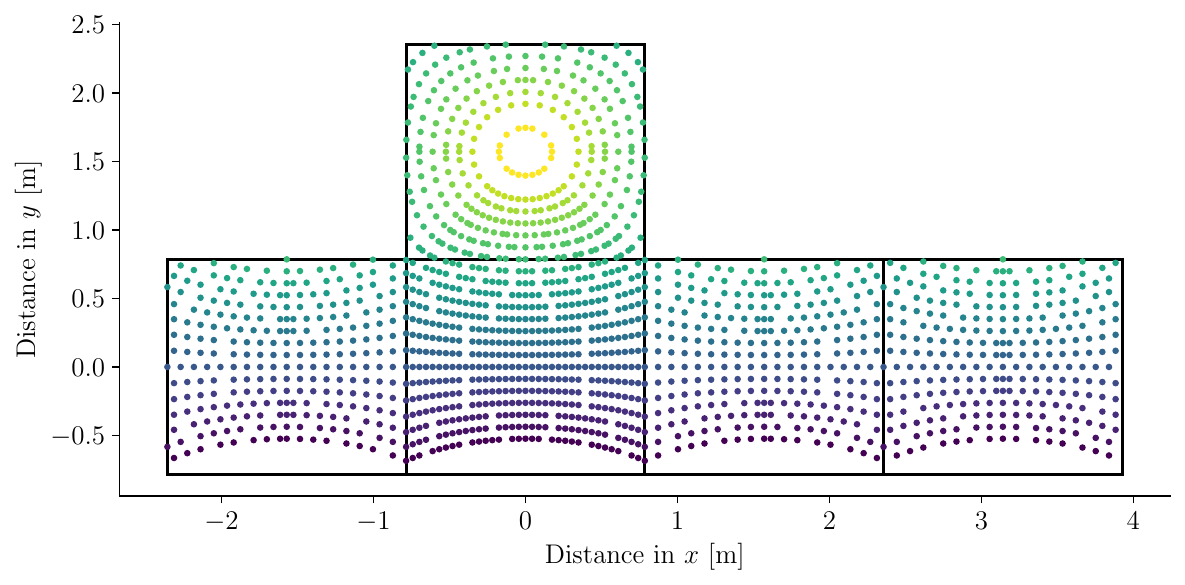}};
   \node [below left,align=center, fill=white, fill opacity=0.8, text opacity= 1.0] at ([yshift=-2.55cm, xshift=-1.86cm]img.north east){\footnotesize{$n=3$}};
   \node [below left,align=center, fill=white, fill opacity=0.8, text opacity= 1.0] at ([yshift=-2.55cm, xshift=-4.05cm]img.north east){\footnotesize{$n=2$}};
   \node [below left,align=center, fill=white, fill opacity=0.8, text opacity= 1.0] at ([yshift=-2.55cm, xshift=-6.25cm]img.north east){\footnotesize{$n=1$}};
   \node [below left,align=center, fill=white, fill opacity=0.8, text opacity= 1.0] at ([yshift=-2.55cm, xshift=-8.45cm]img.north east){\footnotesize{$n=4$}};
   \node [below left,align=center, fill=white, fill opacity=0.8, text opacity= 1.0] at ([yshift=-0.575cm, xshift=-6.25cm]img.north east){\footnotesize{$n=5$}};
  \end{tikzpicture}
} \par\hfill
  \setcounter{subfigure}{1}
  \subfloat[Gnomonic equiangular projected positions of the \acp{IR}. \label{fig:projected_data_sphere}]{%
  \includegraphics[width=0.35\linewidth]{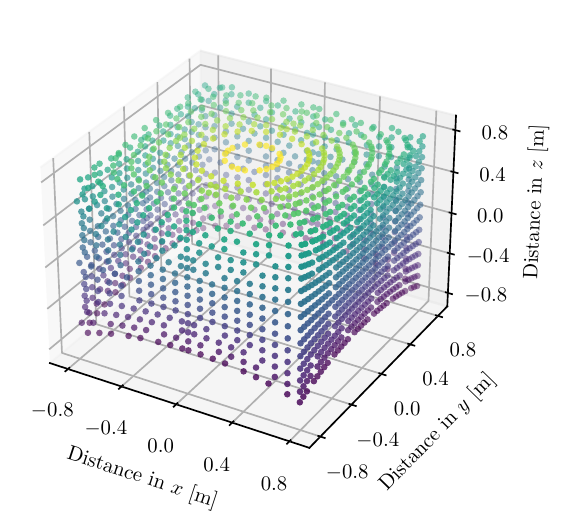}
}
   \hfill
  \setcounter{subfigure}{3}
  \subfloat[Barycentric interpolation of flattened cube projected positions of the \acp{IR}. \label{fig:projected_data_flattened_interpolation}]{
    \begin{tikzpicture}
   \node (img) {\includegraphics[width=0.6\linewidth]{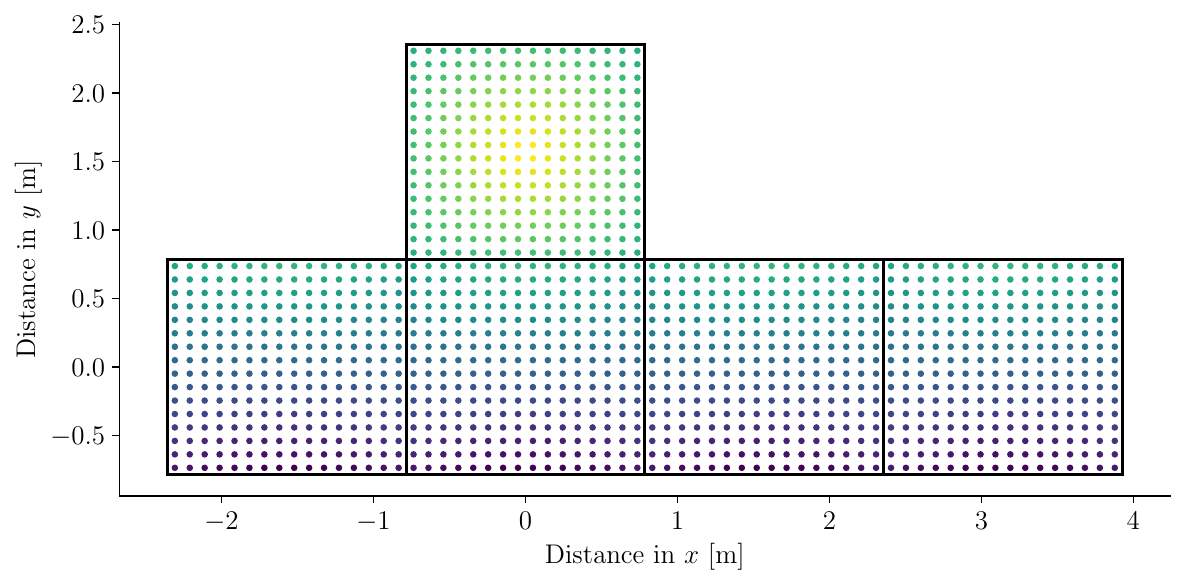}};
   \node [below left,align=center, fill=white, fill opacity=0.8, text opacity= 1.0] at ([yshift=-2.55cm, xshift=-1.86cm]img.north east){\footnotesize{$n=3$}};
   \node [below left,align=center, fill=white, fill opacity=0.8, text opacity= 1.0] at ([yshift=-2.55cm, xshift=-4.05cm]img.north east){\footnotesize{$n=2$}};
   \node [below left,align=center, fill=white, fill opacity=0.8, text opacity= 1.0] at ([yshift=-2.55cm, xshift=-6.25cm]img.north east){\footnotesize{$n=1$}};
   \node [below left,align=center, fill=white, fill opacity=0.8, text opacity= 1.0] at ([yshift=-2.55cm, xshift=-8.45cm]img.north east){\footnotesize{$n=4$}};
   \node [below left,align=center, fill=white, fill opacity=0.8, text opacity= 1.0] at ([yshift=-0.575cm, xshift=-6.25cm]img.north east){\footnotesize{$n=5$}};
  \end{tikzpicture}
} \hfill
 \caption{The four steps that project the \acf{IR} locations for a given \acf{HRTF} in the ARI dataset onto a flattened uniform cube.}
\label{fig:projection_procedure}
\end{figure*}

\subsection{Data pre-processing}
\label{sec:data_pre_processing}
\Acp{SRGAN} have been shown to perform well on the task of upsampling images. The main challenge when it comes to upsampling \acp{HRIR} is that the data occupies an extra dimension in space compared to a \ac{2D} image. Another issue is that \acp{CNN} are designed for applications using uniformly spaced data (such as the pixels in conventional \ac{2D} images). In contrast, the \acp{IR} in an \ac{HRIR} are spaced non-uniformly on the surface of a sphere. In particular, \acp{HRIR} often contain no measurements at lower elevations, and the number of measurements is denser around the horizontal plane.

Various approaches exist for processing non-uniformly distributed spatial data, such as \acp{GNN} \cite{Zhou2020}. However, in order to exploit the vast literature that exists for the upsampling of images, in this study, we apply a pre-processing step to convert the spherical data into a form that a standard \ac{CNN} can process.

Two main steps are required to convert the \ac{HRIR} data into a format that can be exploited by a \ac{CNN} architecture. First, the spherical \ac{HRIR} data needs to be projected onto a \ac{2D} surface to remove the extra dimension (see \cref{sec:gnomonic_equiangular_projection}). Second, an interpolation is utilised to shift the irregularly spaced \acp{IR} onto an evenly spaced Cartesian grid (see \cref{sec:barycentric_interpolation}). This strategy has the advantage of mapping any \ac{HRIR} dataset to the same cartesian grid, and as a result, any dataset can be deployed for training and testing the \ac{SRGAN} \cite{Hogg2023b}. 

In addition to these two steps, the phase and \acp{ITD} were disregarded by taking the \ac{HRIR} into the frequency domain, referred to as the \ac{HRTF}, and only considering the magnitude component from each \ac{IR} in the \ac{HRIR}. These additional simplifying pre-processing steps can be performed because the up-sampled \acp{HRTF} can be effectively reconstructed using a minimum-phase approximation and a simple ITD model \cite{Cuevas-Rodriguez2019}. However, it should be noted that such simplifications could have an impact on certain perceptual features of the \acp{HRTF} \cite{Andreopoulou2022}. For this reason, future advancements in this technique should aim to include phase information when performing the upsampling.

\subsubsection{Gnomonic equiangular projection}
\label{sec:gnomonic_equiangular_projection}
A gnomonic equiangular projection \cite{Jung2019} is used to project the locations of the \acp{IR}, seen in \crefsub{fig:orginal_data_sphere}, in the \ac{HRIR} to a cube, shown in \crefsub{fig:projected_data_sphere}, which can then be flattened as shown in \crefsub{fig:projected_data_flattened}. This process creates five panels where identical local curvilinear coordinates are constructed for each panel \cite{Ronchi1996}. It should be noted that the 6$^\text{th}$ (bottom) panel is removed as it contains no \ac{IR} measurements as the \acp{HRIR} are usually not measured below the listener. More information about this choice can be found later on.

The gnomonic equiangular projection \cite{Jung2019} transforms the location of any point on the sphere into Cartesian coordinates $x$ and $y$  using
\begin{equation}
    x = r\left(\theta-\frac{(n-1)}{2}\pi \right)\;,
\end{equation}
\begin{equation}
    y = r\arctan\left[\tan(\phi)\sec\left(\theta-\frac{(n-1)}{2}\pi \right)\right]\;.
\end{equation}
where $n$ represents the equatorial panel on the horizontal plane that the point on the sphere would map onto, which is determined based on the original elevation and azimuth and can only take the values 1, 2, 3 and 4. However, the location of any point on the top of the sphere, i.e. when $n = 5$, is mapped to Cartesian coordinates $x$ and $y$ using
\begin{equation}
    x = r\arctan\big(\sin(\theta)\cot(\phi)\big)\;,
\end{equation}\begin{equation}
    y = r\arctan\big(-\cos(\theta)\cot(\phi)\big)\;,
\end{equation}
where $\theta$, $\phi$ and $r$ correspond to the azimuth, elevation and radius of each point on the original sphere. More precisely, the azimuth $\theta$ is defined as the angle between the projection of the source direction in the horizontal plane and the front axis in the range of [$\pi$, $-\pi$] going from left to right, and elevation $\phi$ is defined as the angle between the horizontal plane and the position of the source.

\subsubsection{Interaural time difference removal}
\label{sec:ITD_removal}

Due to the anatomy of the head, sounds from different directions will inevitably arrive with different time delays. These delays are referred to as the \acp{ITD}. These delays can be removed and then reconstructed after the upsampling has occurred using a simple \ac{ITD} model. \cite{Andreopoulou2022}.

In this paper, a Kalman filter \cite{Kalman1960} is used to detect the onset of each \ac{IR} in the \acp{HRTF} so that the \ac{ITD} can be removed. This onset detection method is similar to that of \cite{hogg2019, Hogg2021d} and works on the assumption that the amplitude behaviour of the noise floor is predictable and, therefore, if the error in the prediction is large, then the amplitude could not be predicted, and the onset of \ac{IR} has occurred.

The amplitude of the \ac{IR}, $x_n$, for the time index, $n$, is modelled here as a random walk with zero-mean, normally distributed increments such that
\begin{alignat}{2}
   x_n &=  x_{n-1}+w, \quad w \in \mathcal{N}(0,\, \sigma^2_w)\;, 
\end{alignat}
where the amplitude at $n$ deviates from the amplitude at ${n-1}$ with a variance of $\sigma^2_w$, and $x_0$ is defined as zero. Observations of the \ac{IR}, $z_n$, are modelled  conditionally on $x_n$ as
\begin{alignat}{2}
  z_n &= x_n + v, \quad v \in \mathcal{N}(0,\, \sigma^2_v)\;, 
\end{alignat}
where the measurement noise, $v$, in this case, models the uncertainty in the observations. 

The Kalman filter estimates the system's state and then acquires feedback from noisy measurements using a prediction and update step. The predicted amplitude estimate, $\hat{x}_{n|n-1}$, and predicted estimate variance, $P_{n|n-1}$, are given by
\begin{align} 
    \hat{x}_{n|n-1} &= \hat{x}_{n-1|n-1}\;,  \label{eq:1}  \\
    P_{n|n-1} &= P_{n-1|n-1} + \sigma^2_w\;.  \label{eq:2} 
\end{align}
The updated amplitude estimate, $\hat{x}_{n|n}$, and updated estimate variance, $P_{n|n}$, are given by
\begin{align} 
    \hat{x}_{n|n} &= \hat{x}_{n|n-1} + K_n(z_n-\hat{x}_{n|n-1})\;, \label{eq:3}\\ 
    P_{n|n} &= (1 - K_n)^2P_{n|n-1} + K_n^2\sigma^2_v\;. \label{eq:4}
\end{align}
Where the innovation variance, $S_n$, and optimal Kalman gain, $K_n$, are given by
\begin{align}
    S_n &= P_{n|n-1}+\sigma^2_v\;,\\
    K_n &= \frac{P_{n|n-1}}{S_n}\;.
\end{align}
The error between measurement and prediction can, therefore, be calculated as
\begin{equation} \label{eq:12}
    \tilde{y}_{n|n} = z_{n} - \hat{x}_{n|n}\;.
\end{equation}
If this error, $\tilde{y}_{n|n}$, is above a threshold, $\eta$, then that implies that the error is large and the value, $x_n$, could not be predicted. This is indicative of the onset of the \ac{IR}, which is not predictable by the Kalman filter. Therefore, once the onset has been located, the \ac{IR} is trimmed before and after the onset so that all the \acp{IR} in the \ac{HRTF} possess the same delay, thus removing the \ac{ITD}.

\subsubsection{Barycentric interpolation}
\label{sec:barycentric_interpolation}

The gnomonic equiangular projection (shown in \crefsub{fig:projected_data_flattened}) has transformed the \ac{3D} space into a \ac{2D} plane; however, the issue of the measurements being spaced at irregular intervals still remains as the Cartesian points lie along curves. This is a problem as the convolution kernels used by the \ac{CNN} require a uniform grid to function correctly. Therefore, barycentric interpolation \cite{Cuevas-Rodriguez2019} is used to project the data onto a regular Cartesian grid. For simplicity, the barycentric interpolation is performed on the sphere of \acp{IR} before the \acp{IR} are mapped using the gnomonic projection.     

In previous work using barycentric interpolation \cite{Cuevas-Rodriguez2019}, the three nearest measurement points to the interpolated point are calculated. However, suppose the measurement points are not evenly spaced, which is the case here. In that case, this can lead to the issue of the three selected measurement points not forming a spherical triangle around the point to be interpolated. Therefore, to solve this problem, we take the three closest measurement points, forming a spherical triangle around the point to be interpolated. This is similar to \cite{Gamper2013}, which proposes a barycentric interpolation among the \acp{HRIR} at the points of a 3D tetrahedron conformed by four measurement points which surround the point to be interpolated.

First, in order to perform barycentric interpolation, it is necessary to find the three closest points that form a spherical triangle ($P_1$, $P_2$ and $P_3$) around the interpolated point ($P_i$), where a spherical triangle can be defined as a curved surface on a sphere which is bounded by the arcs of three great circles. Second, the barycentric coordinates $\alpha$, $\beta$, and $\gamma$ need to be calculated. These coordinates represent the ratio of the areas of the three smaller triangles ($P_iP_2P_3$, $P_1P_iP_3$, and $P_1P_2P_i$) relative to the larger triangle ($P_1P_2P_3$), such that $\alpha + \beta + \gamma = 1$. The coefficients $\alpha$, $\beta$, and $\gamma$ correspond to weights applied to the \acp{IR} at points $P_1$, $P_2$, and $P_3$, respectively. Ultimately, these coefficients will be used to find the interpolated HRIR for point $P_i$.  

In \cite{Cuevas-Rodriguez2019}, elevation ($\phi$) and azimuth ($\theta$) are treated as Cartesian coordinates, and the following formulas are used to find the ratio of the areas
\begin{equation}
\resizebox{0.89\hsize}{!}{$
    \alpha = \frac
    {(\phi^{P_2}-\phi^{P_3})(\theta^{P_i}-\theta^{P_3})+(\theta^{P_3}-\theta^{P_2})(\phi^{P_i}-\phi^{P_3})}
    {(\phi^{P_2}-\phi^{P_3})(\theta^{P_1}-\theta^{P_3})+(\theta^{P_3}-\theta^{P_2})(\phi^{P_1}-\phi^{P_3})} 
$\;,}
\label{eq:alpha}
\end{equation}
\begin{equation}
\resizebox{0.89\hsize}{!}{$
    \beta = \frac
    {(\phi^{P_3}-\phi^{P_1})(\theta^{P_i}-\theta^{P_3})+(\theta^{P_1}-\theta^{P_3})(\phi^{P_i}-\phi^{P_3})}
    {(\phi^{P_2}-\phi^{P_3})(\theta^{P_1}-\theta^{P_3})+(\theta^{P_3}-\theta^{P_2})(\phi^{P_1}-\phi^{P_3})} 
$\;,}
\label{eq:beta}
\end{equation}
\begin{equation}
    \gamma = 1 - \alpha - \beta\;.
    \label{eq:gamma}
\end{equation}
Then, in order to treat elevation and azimuth as spherical coordinates rather than Cartesian coordinates, L'Huilier's Theorem \cite{Zwillinger2018} is used. L'Huilier's Theorem states that the surface area of a spherical triangle is given by $A = r^2E$, where $r$ is the radius of the sphere, and $E$ is the excess angle. The excess angle, $E$, is defined by 
\begin{equation}
\resizebox{0.89\hsize}{!}{$
    E = 4\times\arctan\Vast(\sqrt{
    \begin{aligned}
      \tan\big(\frac{1}{2}s\big)&\tan\big(\frac{1}{2}(s-a)\big)\\
      \times&\tan\big(\frac{1}{2}(s-b)\big)\tan\big(\frac{1}{2}(s-c)\big)
    \end{aligned}
    }$\Vast)\,,}
    \label{eq:lhuilier}
\end{equation}
where $a$, $b$, $c$ represent the side arc lengths of the triangle calculated using the haversine distance between two points on a sphere and $s=(a+b+c)/2$. The equations \cref{eq:alpha,eq:beta,eq:gamma} can then be modified using \cref{eq:lhuilier} to use spherical coordinates to obtain the new weights
\begin{equation}
    \alpha = \frac
    {E^{P_iP_2P_3}}
    {E^{P_1P_2P_3}}\;, \quad
    \beta = \frac
    {E^{P_1P_iP_3}}
    {E^{P_1P_2P_3}}\;, \quad
    \gamma = 1 - \alpha - \beta\;. 
    \label{eq:pherical_coordinate}
\end{equation}

\subsubsection{Magnitude spectrum extraction}
\label{sec:magnitude_extraction}

After removing the \ac{ITD}, the \acp{HRIR} are interpolated for each point of interest, $P_i$, using the barycentric coordinates $\alpha$, $\beta$, and $\gamma$ (see \cref{sec:barycentric_interpolation}) along with 
\begin{equation}
    \text{HRIR}^{P_i} = \alpha \text{HRIR}^{P_1} + \beta \text{HRIR}^{P_2} + \gamma \text{HRIR}^{P_3}\;.
\end{equation}
Following interpolation, the \ac{HRIR} is transformed into the \ac{HRTF} via the \ac{DFT}. The magnitude of the \ac{HRTF} is then used as an input to the \acp{GAN}.

\begin{figure}[tb]
\centering

  \subfloat[\label{fig:padding_procedure_cube}]{
  \foreach \myPsi in {135}{
    \tdplotsetmaincoords{70}{\myPsi}
    \begin{tikzpicture}[scale=0.25]
        \clip (-7,-6) rectangle (7,6);
        \begin{scope}[tdplot_main_coords]
            \draw[step=2cm,canvas is yx plane at z=4, pattern=north west lines, pattern color=c4] (-4,2) rectangle (4,4); 
            \draw[step=2cm,canvas is xz plane at y=4, pattern=north east lines, pattern color=c1] (2,-4) rectangle (4,4);
            \draw[step=2cm,canvas is yz plane at x=4, pattern=crosshatch, pattern color=c3] (-4,-4) rectangle (4,4);
            \draw[step=2cm,canvas is yz plane at x=4] (-4.01,-4.01) grid (4,4);
            \draw[step=2cm,canvas is xz plane at y=4] (-4.01,-4.01) grid (4,4);
            \draw[step=2cm,canvas is yx plane at z=4] (-4.01,-4.01) grid (4,4);
        \end{scope}
    \end{tikzpicture}}
} \hspace{0.4cm}
  \subfloat[ \label{fig:padding_procedure_flattened}]{%
  \begin{tikzpicture}[scale=0.25]
        \clip (-7,-7) rectangle (7,7);
        \draw[step=2cm, pattern=crosshatch, pattern color=c3] (-6,2) rectangle (2,-6); 
        % \draw[step=2cm, pattern=north east lines, pattern color=c1] (2,2) -- (2,-6) -- (4,-6)  -- (4,4) -- cycle;
        % \draw[step=2cm, pattern=north west lines, pattern color=c4] (2,2) -- (-6,2) -- (-6,4)  -- (4,4) -- cycle; 
        \draw[step=2cm, pattern=north east lines, pattern color=c1] (2,2) -- (2,-6) -- (4,-6)  -- (4,2) -- cycle;
        \draw[step=2cm, pattern=north west lines, pattern color=c4] (4,2) -- (-6,2) -- (-6,4)  -- (4,4) -- cycle; 
        \draw[step=2cm,color=black] (-6.01,-6.01) grid (4.0,4.0);
    \end{tikzpicture}
}
 \caption{Each face of the gnomonic equiangular projection (in green) is padded with data from the adjacent faces (in red). This is displayed both for the 3D cube (a) and for the flattened 2D surface (b). In the corner, the value is ambiguous, therefore, values are taken from the top panel \cite{Weyn2020}.}
\label{fig:padding_procedure}
\end{figure}
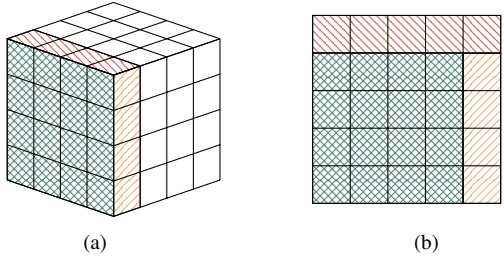
\begin{figure*}[t!]
	\centering
  \subfloat[Discriminator network $D$.\label{fig:discriminator_network}]{%
        \begin{tikzpicture}[rotate=90,transform shape,
            node distance=4mm and 3.5mm,
            start chain=going below,
         mynode/.style = {
                draw=gray, rectangle, align=center, text width=1.9cm,
                font=\small, inner sep=1.0ex, outer sep=0pt, minimum width=2.3cm,
                on chain},
         every join/.style = arrow,
         arrow/.style = {very thick,-stealth}
                            ] 

        \node (n1)  [mynode, fill=c9] {\figfont Input};
        \node (n2)  [mynode,below= of n1, yshift=2mm, fill=c1] {\figfont GnomicProjConv};
        \node (n3)  [mynode,below= of n2, yshift=2mm, fill=c6, text=white] {\figfont Leaky ReLU};

        \node (n4a)  [mynode,below= of n3, yshift=1.1mm, minimum height=1.98cm,minimum width=2.5cm, fill=c5] {};
        \node (n4)  [mynode,below= of n3, yshift=0mm, fill=c1] {\figfont GnomicProjConv};
        \node (n5)  [mynode,below= of n4, yshift=3mm, fill=c7] {\figfont Batch Norm};
        \node (n6)  [mynode,below= of n5, yshift=3mm, fill=c6, text=white] {\figfont Leaky ReLU};

        \node (n7a)  [mynode,below= of n6, yshift=-0.7mm, minimum height=0.8cm,minimum width=2.5cm, fill=c5] {};
        \node (n7)  [mynode,below= of n6, yshift=-2mm, minimum height=0.15cm, minimum width=2.3cm, inner sep=0ex, fill=c1] {};
        \node (n8)  [mynode,below= of n7, yshift=3.5mm, minimum height=0.15cm, minimum width=2.3cm, inner sep=0ex, fill=c7] {};
        \node (n9)  [mynode,below= of n8, yshift=3.5mm, minimum height=0.15cm, minimum width=2.3cm, inner sep=0ex, fill=c6] {};
        
        \node (n10a)  [mynode,below= of n9, yshift=-0.7mm, minimum height=0.8cm,minimum width=2.5cm, fill=c5] {};
        \node (n10)  [mynode,below= of n9, yshift=-2mm, minimum height=0.15cm, minimum width=2.3cm, inner sep=0ex, fill=c1] {};
        \node (n11)  [mynode,below= of n10, yshift=3.5mm, minimum height=0.15cm, minimum width=2.3cm, inner sep=0ex, fill=c7] {};
        \node (n12)  [mynode,below= of n11, yshift=3.5mm, minimum height=0.15cm, minimum width=2.3cm, inner sep=0ex, fill=c6] {};
        
        \node (n13a)  [mynode,below= of n12, yshift=-0.7mm, minimum height=0.8cm,minimum width=2.5cm, fill=c5] {};
        \node (n13)  [mynode,below= of n12, yshift=-2mm, minimum height=0.15cm, minimum width=2.3cm, inner sep=0ex, fill=c1] {};
        \node (n14)  [mynode,below= of n13, yshift=3.5mm, minimum height=0.15cm, minimum width=2.3cm, inner sep=0ex, fill=c7] {};
        \node (n15)  [mynode,below= of n14, yshift=3.5mm, minimum height=0.15cm, minimum width=2.3cm, inner sep=0ex, fill=c6] {};
        
        \node (n16a)  [mynode,below= of n15, yshift=-0.7mm, minimum height=0.8cm,minimum width=2.5cm, fill=c5] {};
        \node (n16)  [mynode,below= of n15, yshift=-2mm, minimum height=0.15cm, minimum width=2.3cm, inner sep=0ex, fill=c1] {};
        \node (n17)  [mynode,below= of n16, yshift=3.5mm, minimum height=0.15cm, minimum width=2.3cm, inner sep=0ex, fill=c7] {};
        \node (n18)  [mynode,below= of n17, yshift=3.5mm, minimum height=0.15cm, minimum width=2.3cm, inner sep=0ex, fill=c6] {};
        
        \node (n19a)  [mynode,below= of n18, yshift=-0.7mm, minimum height=0.8cm,minimum width=2.5cm, fill=c5] {};
        \node (n19)  [mynode,below= of n18, yshift=-2mm, minimum height=0.15cm, minimum width=2.3cm, inner sep=0ex, fill=c1] {};
        \node (n20)  [mynode,below= of n19, yshift=3.5mm, minimum height=0.15cm, minimum width=2.3cm, inner sep=0ex, fill=c7] {};
        \node (n21)  [mynode,below= of n20, yshift=3.5mm, minimum height=0.15cm, minimum width=2.3cm, inner sep=0ex, fill=c6] {};
        
        \node (n22a)  [mynode,below= of n21, yshift=-0.7mm, minimum height=0.8cm,minimum width=2.5cm, fill=c5] {};
        \node (n22)  [mynode,below= of n21, yshift=-2mm, minimum height=0.15cm, minimum width=2.3cm, inner sep=0ex, fill=c1] {};
        \node (n23)  [mynode,below= of n22, yshift=3.5mm, minimum height=0.15cm, minimum width=2.3cm, inner sep=0ex, fill=c7] {};
        \node (n24)  [mynode,below= of n23, yshift=3.5mm, minimum height=0.15cm, minimum width=2.3cm, inner sep=0ex, fill=c6] {};
        
        \node (n25)  [mynode,below= of n24, yshift=-1mm, fill=c8] {\figfont Dense (512)};
        \node (n26)  [mynode,below= of n25, yshift=2mm, fill=c6, text=white] {\figfont Leaky ReLU};
        \node (n27)  [mynode,below= of n26, yshift=2mm, fill=c8] {\figfont Dense (1)};
        \node (n28)  [mynode,below= of n27, yshift=2mm, fill=black, text=white] {\figfont Sigmoid};

        \node (n29) [below= of n28, yshift=-5mm] {};

       \begin{scope}[on background layer]
         \draw[arrow, gray, line width=2mm] (n1) -- (n29);
       \end{scope}

       \node[text width=0cm, rotate=-90, xshift=-0.55cm, yshift=0.3cm] at (n2.east) {\footnotesize k3n64s1};
       \node[text width=0cm, rotate=-90, xshift=-0.55cm, yshift=0.3cm] at (n4.east) {\footnotesize k3n64s1};
       
       \node[text width=0cm, rotate=-90, xshift=-0.55cm, yshift=0.3cm] at (n7.east) {\footnotesize k3n128s1};
       \node[text width=0cm, rotate=-90, xshift=-0.55cm, yshift=0.6cm] at (n10.east) {\footnotesize k3n128s2};

       \node[text width=0cm, rotate=-90, xshift=-0.55cm, yshift=0.3cm] at (n13.east) {\footnotesize k3n256s1};
       \node[text width=0cm, rotate=-90, xshift=-0.55cm, yshift=0.6cm] at (n16.east) {\footnotesize k3n256s2};

       \node[text width=0cm, rotate=-90, xshift=-0.55cm, yshift=0.3cm] at (n19.east) {\footnotesize k3n512s1};
       \node[text width=0cm, rotate=-90, xshift=-0.55cm, yshift=0.6cm] at (n22.east) {\footnotesize k3n512s2};

       % Add white space
       \node[xshift=-2.5cm] at (n1.east) {\footnotesize };

       \end{tikzpicture}  
}
\vspace{0.1cm}
  \subfloat[Generator network $G$.\label{fig:generator_network}]{
        \begin{tikzpicture}[rotate=90,transform shape,
            node distance=4mm and 3.5mm,
            start chain=going below,
         mynode/.style = {
                draw=gray, rectangle, align=center, text width=1.9cm,
                font=\small, inner sep=1.0ex, outer sep=0pt, minimum width=2.3cm,
                on chain},
         every join/.style = arrow,
         arrow/.style = {very thick,-stealth}
                            ] 
        
        \node (n1)  [mynode, fill=c9] {\figfont Input};
        \node (n2)  [mynode,below= of n1, yshift=2mm, fill=c1] {\figfont GnomicProjConv};
        \node (n3)  [mynode,below= of n2, yshift=2mm, fill=c10, text=white] {\figfont PReLU};

        \node (n4a)  [mynode,below= of n3, yshift=-0.9mm, minimum height=3.06cm,minimum width=2.5cm, fill=c5] {};
        \node (n4)  [mynode,below= of n3, yshift=-2mm, fill=c1] {\figfont GnomicProjConv};
        \node (n5)  [mynode,below= of n4, yshift=3mm, fill=c7] {\figfont Batch Norm};
        \node (n6)  [mynode,below= of n5, yshift=3mm, fill=c10, text=white] {\figfont PReLU};
        \node (n7)  [mynode,below= of n6, yshift=3mm, minimum height=0.2cm, minimum width=2.3cm, inner sep=0ex, fill=c1] {};
        \node (n8)  [mynode,below= of n7, yshift=3.5mm, minimum height=0.2cm, minimum width=2.3cm, inner sep=0ex, fill=c7] {};
        \node (n9)  [mynode,below= of n8, yshift=3mm, fill=c4] {\figfont Elementwise Sum};

        \node (n10a)  [mynode,below= of n9, yshift=-0.7mm, minimum height=1.4cm,minimum width=2.5cm, fill=c5] {};
        \node (n10)  [mynode,below= of n9, yshift=-2mm, minimum height=0.15cm, minimum width=2.3cm, inner sep=0ex, fill=c1] {};
        \node (n11)  [mynode,below= of n10, yshift=3.5mm, minimum height=0.15cm, minimum width=2.3cm, inner sep=0ex, fill=c7] {};
        \node (n12)  [mynode,below= of n11, yshift=3.5mm, minimum height=0.15cm, minimum width=2.3cm, inner sep=0ex, fill=c10] {};
        \node (n13)  [mynode,below= of n12, yshift=3.5mm, minimum height=0.15cm, minimum width=2.3cm, inner sep=0ex, fill=c1] {};
        \node (n14)  [mynode,below= of n13, yshift=3.5mm, minimum height=0.15cm, minimum width=2.3cm, inner sep=0ex, fill=c7] {};
        \node (n15)  [mynode,below= of n14, yshift=3.5mm, minimum height=0.15cm, minimum width=2.3cm, inner sep=0ex, fill=c4] {};

        \node (n16a)  [mynode,below= of n15, yshift=-4.7mm, minimum height=1.4cm,minimum width=2.5cm, fill=c5] {};
        \node (n16)  [mynode,below= of n15, yshift=-6mm, minimum height=0.15cm, minimum width=2.3cm, inner sep=0ex, fill=c1] {};
        \node (n17)  [mynode,below= of n16, yshift=3.5mm, minimum height=0.15cm, minimum width=2.3cm, inner sep=0ex, fill=c7] {};
        \node (n18)  [mynode,below= of n17, yshift=3.5mm, minimum height=0.15cm, minimum width=2.3cm, inner sep=0ex, fill=c10] {};
        \node (n19)  [mynode,below= of n18, yshift=3.5mm, minimum height=0.15cm, minimum width=2.3cm, inner sep=0ex, fill=c1] {};
        \node (n20)  [mynode,below= of n19, yshift=3.5mm, minimum height=0.15cm, minimum width=2.3cm, inner sep=0ex, fill=c7] {};
        \node (n21)  [mynode,below= of n20, yshift=3.5mm, minimum height=0.15cm, minimum width=2.3cm, inner sep=0ex, fill=c4] {};

        \node (n22)  [mynode,below= of n21, yshift=0mm, fill=c1] {\figfont GnomicProjConv};
        \node (n23)  [mynode,below= of n22, yshift=2mm, fill=c7] {\figfont Batch Norm};
        \node (n24)  [mynode,below= of n23, yshift=2mm, fill=c4] {\figfont Elementwise Sum};

        \node (n25a)  [mynode,below= of n24, yshift=1.1mm, minimum height=1.98cm,minimum width=2.5cm, fill=c5] {};
        \node (n25)  [mynode,below= of n24, yshift=0mm, fill=c1] {\figfont GnomicProjConv};
        \node (n26)  [mynode,below= of n25, yshift=3mm, fill=c12] {\figfont Pixel Shuffle $\times 2$};
        \node (n27)  [mynode,below= of n26, yshift=3mm, fill=c10, text=white] {\figfont PReLU};

        \node (n28a)  [mynode,below= of n27, yshift=0.3mm, minimum height=0.8cm,minimum width=2.5cm, fill=c5] {};
        \node (n28)  [mynode,below= of n27, yshift=-1mm, minimum height=0.15cm, minimum width=2.3cm, inner sep=0ex, fill=c1] {};
        \node (n29)  [mynode,below= of n28, yshift=3.5mm, minimum height=0.15cm, minimum width=2.3cm, inner sep=0ex, fill=c12] {};
        \node (n30)  [mynode,below= of n29, yshift=3.5mm, minimum height=0.15cm, minimum width=2.3cm, inner sep=0ex, fill=c10] {};

        \node (n31)  [mynode,below= of n30, yshift=0mm, fill=c1] {\figfont GnomicProjConv};
        \node (n32)  [mynode,below= of n31, yshift=2mm, fill=c11, text=white] {\figfont Softplus};

        \node (n33) [below= of n32, yshift=-5mm] {};

       \begin{scope}[on background layer]
         \draw[arrow, gray, line width=2mm] (n1) -- (n33);
       \end{scope}

         \coordinate (a1) at ([yshift=0.27cm]n4.north);
         \coordinate (a2) at ([yshift=0.27cm, xshift=-1.6cm]n4.north);
         \coordinate (a3) at ([xshift=-1.6cm, yshift=0.25cm]n9.south);
         \coordinate (a4) at ([xshift=-1.15cm, yshift=0.25cm]n9.south);
         \draw[arrow, gray, line width=0.5mm] (a1) -- (a2) -- (a3) -- (a4);

         \coordinate (a5) at ([yshift=0.3cm]n10.north);
         \coordinate (a6) at ([yshift=0.3cm, xshift=-1.6cm]n10.north);
         \coordinate (a7) at ([xshift=-1.6cm, yshift=0.075cm]n15.south);
         \coordinate (a8) at ([xshift=-1.15cm, yshift=0.075cm]n15.south);
         \draw[arrow, gray, line width=0.5mm] (a5) -- (a6) -- (a7) -- (a8);

         \coordinate (a9) at ([yshift=0.3cm]n16.north);
         \coordinate (a10) at ([yshift=0.3cm, xshift=-1.6cm]n16.north);
         \coordinate (a11) at ([xshift=-1.6cm, yshift=0.075cm]n21.south);
         \coordinate (a12) at ([xshift=-1.15cm, yshift=0.075cm]n21.south);
         \draw[arrow, gray, line width=0.5mm] (a9) -- (a10) -- (a11) -- (a12);

         \coordinate (a13) at ([yshift=0.45cm]n4.north);
         \coordinate (a14) at ([yshift=0.45cm, xshift=-1.79cm]n4.north);
         \coordinate (a15) at ([xshift=-1.79cm, yshift=0.25cm]n24.south);
         \coordinate (a16) at ([xshift=-1.15cm, yshift=0.25cm]n24.south);
         \draw[arrow, gray, line width=0.5mm] (a13) -- (a14) -- (a15) -- (a16);

         \draw [decorate,decoration={brace,amplitude=10pt},rotate=180, gray] ([xshift=-1.68cm]n4.north) -- ([xshift=-1.68cm]n21.south) node [black,midway,xshift=-0.6cm, rotate=90] {\footnotesize $B$ residual blocks (3 shown)};

         \draw [decorate,decoration={brace,amplitude=10pt},rotate=180, gray] ([xshift=-1.68cm]n25.north) -- ([xshift=-1.68cm]n30.south) node [black,midway,xshift=-0.6cm, rotate=90] {\footnotesize $R$ upsampling blocks (2 shown)};

         \node[text width=0cm, rotate=-90, xshift=-0.55cm, yshift=0.3cm] at (n2.east) {\footnotesize k3n512s1};
         \node[text width=0cm, rotate=-90, xshift=-0.55cm, yshift=0.3cm] at (n4.east) {\footnotesize k3n512s1};
         \node[text width=0cm, rotate=-90, xshift=-0.55cm, yshift=0.3cm] at (n7.east) {\footnotesize k3n512s1};

         \node[text width=0cm, rotate=-90, xshift=-0.55cm, yshift=0.3cm] at (n22.east) {\footnotesize k3n512s1};
         \node[text width=0cm, rotate=-90, xshift=-0.55cm, yshift=0.3cm] at (n25.east) {\footnotesize k3n4096s1};
         \node[text width=0cm, rotate=-90, xshift=-0.55cm, yshift=0.3cm] at (n31.east) {\footnotesize k3n128s1};

        % Add white space
        \node[xshift=-3cm] at (n1.east) {\footnotesize };
       
        \end{tikzpicture}
} 
\caption{The architecture of the discriminator and generator networks, where each convolutional layer contains $k$ kernels, $n$ feature layers, and $s$ stride. Acronyms: \acf{LReLU}, \acf{PReLU}.}
	\label{fig:GAN_architecture}
\end{figure*}
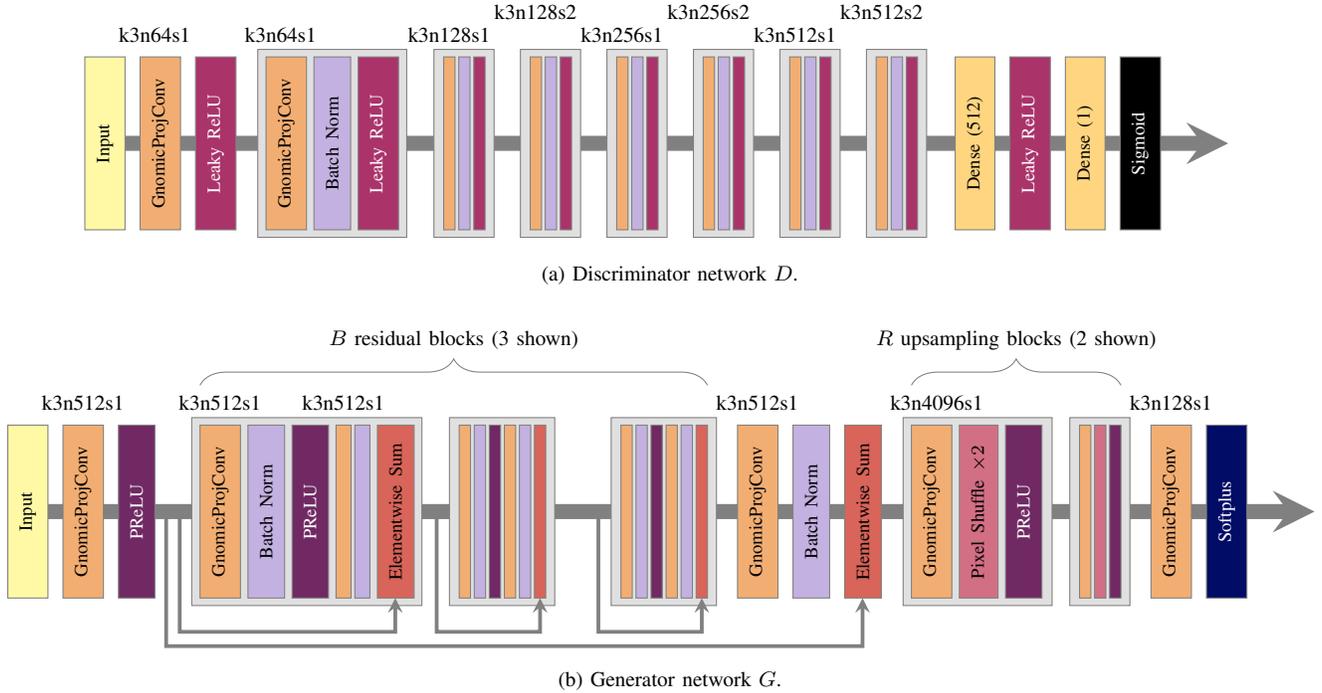

\subsection{GAN architecture}
\label{sec:GAN_architecture}

In this work, a \ac{GAN} architecture \cite{Goodfellow2014} is used to generate high-resolution \acp{HRTF}, $H_{\text{HR}}$, from their low-resolution counterparts, $H_{\text{LR}}$. This is achieved through a supervised learning approach where the network has access to the high-resolution $H_{\text{HR}}$ target during training. 

To generate the low-resolution $H_{\text{LR}}$, the high-resolution $H_{\text{HR}}$ is downsampled by a factor of $r$. The dimensions of $H_{\text{HR}}$ are $B\times X\times W\times H$, where $B$, $X$, $W$ and $H$ represent respectively the number of frequency bins, the number of cube sphere panels, the height of each cube sphere panel, and the width of each cube sphere panel. Therefore, after downsampling, considering that we are using $X=5$ panels for this implementation, the dimensions of $H_\text{LR}$ are $B\times 5\times \frac{W}{r}\times \frac{H}{r}$. It is important to note that the downsampling is only spatial; therefore, only $W$ and $H$ are downsampled by $r$. In contrast, frequency resolution and the number of cube sphere panels remain the same.

The \ac{GAN} architecture that is exploited in this paper is similar to that of \cite{Ledig2017} and is commonly referred to as \ac{SRGAN}. The \ac{SRGAN} architecture relies on two networks competing in a minimax game, with the generator consisting of residual layers followed by upsampling layers with a low-resolution image as the input and the discriminator taking a high-resolution input and then performing a series of convolutions. This network was chosen as the foundation for this work as it has proven successful at a diverse range of super-resolution tasks. The novelty in this work is that the \ac{2D} convolutional layers are adapted to be able to handle the gnomonic equiangular projection input data where one set of weights is learned for the equatorial panels ($n = 1 \text{ to }  4$) and a separate set of weights is learned for the top panel ($n=5$), i.e. the convolution takes an input of size $B\times 1\times \frac{W}{r}\times \frac{H}{r}$ and learns two sets of weights, one for the top panel and a shared weight set for the equatorial panels. This differs from previous approaches that learn a single set of convolutional weights for all panels of the gnomonic equiangular projection, e.g. \cite{Weyn2020}. In addition, a novel gnomonic equiangular projection padding layer is added before each convolutional layer in the discriminator and generator models; this layer pads each panel in the gnomonic equiangular projection based on its adjacent panels, as shown in \cref{fig:padding_procedure}. In the corner, the value is ambiguous, therefore, values are taken from the top panel (see \crefsub{fig:padding_procedure_flattened}). The GnomonicProjConv layer is defined as the gnomonic equiangular projection padding layer followed by the adapted convolutional layer. 

In cases where there is no adjacent face, i.e. the lower edge of the equatorial panels, that edge is just padded by repeating the values that are closest to that edge. Note that because the padding is added repeatedly throughout the generator and discriminator networks, the networks can learn from points that stretch around the corners of the gnomonic equiangular projection. 

\acp{GAN} consist of a discriminator network $D$, shown in \crefsub{fig:discriminator_network}, which is optimised alongside a generator network $G$, shown in \crefsub{fig:generator_network}, in an alternating manner to find a solution to the adversarial minimax problem 
\begin{equation}
    \begin{aligned}
          \min_{G} \: \max_{D} \:\: &\mathbb{E}_{H_{\text{HR}}\sim p_{\text{train}}(H_{\text{HR}})}[\log{D(H_{\text{HR}})}] \\
         + &\mathbb{E}_{H_{\text{LR}}\sim p_{\text{G}}(H_{\text{LR}})}[\log{(1 - D(G(H_{\text{LR}})))}]\;.
    \end{aligned}
\end{equation}

\subsubsection{Generator network}
In this work, network $G$ aims to generate high-resolution \acp{HRTF} from low-resolution \ac{HRTF} inputs. The network $G$ consists of $B$ identical residual blocks, each containing two convolutional layers. A batch normalisation layer follows each of these convolutional layers. These batch normalisation layers are followed by a \ac{PReLU} activation layer \cite{He2015a} after the first batch normalisation and an element-wise sum after the second batch normalisation. These element-wise sum units function as an additive residual (skip) connection. 

To increase the \ac{HRTF} 's resolution, $R$ upsampling blocks are added after. Each block has an upsampling factor of 2; therefore, the number of needed blocks, $R$, is related to the downsampling factor, $r$, using $r = 2^R$. The spatial upsampling is performed via a standard pixel shuffle operation, which compresses channels and expands spatial extent by rearranging pixels. This is mathematically equivalent to, but more computationally efficient than, a transposed convolution.

Another convolutional layer then follows these upsampling blocks before a final activation layer. The primary requirement of the activation layer is that the output is constrained to be positive, as the magnitude responses contained in an \ac{HRTF} cannot be negative. There are multiple candidates for this, such as ReLU and Sigmoid. However, in this work, a softplus activation was selected as it is smoother near the origin than a ReLU and has shown better stabilisation and performance properties  \cite{ Zheng2015, Dubey2022a}.

\subsubsection{Discriminator network}
In this work, network $D$ aims to discriminate whether an \ac{HRTF} is real or generated by the network $G$. The network $D$ consists of eight convolutional layers that are immediately followed by batch normalisation with the exception of the first layer. 

Two dense layers finally follow these convolutional layers, and then a Sigmoid activation function provides the discriminator network's output. Apart from the last layer, a leaky \ac{ReLU} is used as the activation function throughout, just as in \cite{Radford2015}. The Leaky \ac{ReLU} activation function is similar to that of the \ac{PReLU} activation function in that both of them are defined as
\begin{equation}
f(x) = \max(ax, x)\;, 
\end{equation}
but in the case of the leaky \ac{ReLU} activation function, $a$ is a hyper-parameter that is set prior to training, while for \ac{PReLU} $a$ is a parameter that is learned during training \cite{He2015a}.

\subsection{Cost functions}
The total loss function $l^\text{US}$ used in the generator network is key to its performance. This function is a weighted sum of the content loss $l_\text{C}^\text{US}$, which compares the upsampled generator output to the high-resolution ground truth, with an adversarial loss $l_\text{A}^\text{US}$, which measures how frequently the generator successfully fools the discriminator network. Therefore, $l^\text{US}$ is defined as
\begin{equation}
    l^\text{US} = \lambda_\text{C}\times l_\text{C}^\text{US} + \lambda_\text{A}\times l_\text{A}^\text{US}\;,
\end{equation}
where multipliers $\lambda_\text{C}$ and $\lambda_\text{A}$ represent the weight assigned to $l_\text{C}^\text{US}$ and $l_\text{A}^\text{US}$.

\subsubsection{Content loss}
The content loss, $l_\text{C}^\text{US}$, is the combination of the \ac{LSD} metric and the \ac{ILD} metric defined as
\begin{equation}
    l_\text{C}^\text{US}= \text{LSD}+\text{ILD}\;.
\end{equation}
The \ac{LSD} metric \cite{Gutierrez-Parera2022} is used in order to score the difference between the target spectrum $H_\text{HR}$ and the generated spectrum $H_\text{US}$.
% as it is more relevant for the audio context than more traditional content loss functions, such as mean squared error.
\begin{equation}
    \resizebox{0.89\hsize}{!}{$
    \text{LSD} = \frac{1}{N} \sum^{N}_{n=1} \sqrt{\frac{1}{B} \sum^{B}_{b=1} \left(20\log_{10}\dfrac{|H_\text{HR}(f_b, x_n)|}{|H_\text{US}(f_b, x_n)|}\right)^2}
    $\;,}
    \label{eq:SD}
\end{equation}
where $|H_\text{HR}(f_b, x_n)|$ and $|H_\text{US}(f_b, x_n)|$ represent the magnitude responses of the high-resolution and up-sampled \ac{HRTF} sets, $B$ is the number of frequency bins in the \ac{HRTF}, $N$ is the number of locations, $f_b$ is the frequency, and $x_n$ is the location. 

The \ac{ILD} metric \cite{McKenzie2019, Engel2022} is defined as
\begin{equation}
    \resizebox{0.89\hsize}{!}{$
    \begin{aligned}
    \text{ILD} = \frac{1}{N}\sum_{n=1}^{N}\frac{1}{B}\sum_{b=1}^{B}
    &\Bigg|\Bigg(20\log_{10}\frac{|H_\text{HR}^\text{Left}(f_{b},x_{n})|}{|H_\text{HR}^\text{Right}(f_{b},x_{n})|}\Bigg) \\ &- \Bigg(20\log_{10}\frac{|H_\text{US}^\text{Left}(f_{b},x_{n})|}{|H_\text{US}^\text{Right}(f_{b},x_{n})|}\Bigg)\Bigg|
    \end{aligned}$\;,}
    \label{eq:LSD}
\end{equation}
where $|H^\text{Left}(f_b, x_n)|$ and $|H^\text{Right}(f_b, x_n)|$ represent the magnitude responses of the left and right ear, respectively. 

The \ac{LSD} and \ac{ILD} metrics then are both z-score normalised where the mean and standard deviation for both the \ac{ILD} and \ac{LSD} were calculated by comparing each \ac{HRTF} in the training set to every other \ac{HRTF}. The \ac{LSD} and \ac{ILD} are then summed to form the content loss function, $l_\text{C}^\text{US}$. This normalisation is to avoid either the \ac{LSD} or \ac{ILD} dominating the gradients during backpropagation if its loss is significantly greater.

% This choice of cost function means that the low-resolution inputs cannot be normalised, as the \ac{LSD} metric relies on \ac{HRTF} magnitudes that have not been normalised.

\subsubsection{Adversarial loss}
The original GAN loss is used as the adversarial loss component of the total loss outlined in \cite{Goodfellow2014}, which relates the generator network's training to the discriminator's output. The adversarial loss is defined over all training samples, $M$, as the binary cross-entropy loss
\begin{equation}
\resizebox{0.89\hsize}{!}{$
\begin{aligned}
    l_\text{A}^\text{US} = -\frac{1}{M}\bigg[\sum^{M}_{m=1}&\Big(y_m\log\big(D(G(H^m_\text{LR}))\big) \\
    &+  (1-y_m)\log\big(1-D(G(H^m_\text{LR}))\big) \Big)\bigg]
\end{aligned}$\;,}
\label{eq:BCE}
\end{equation}

\subsection{Data post-processing}

To carry out some of the evaluations described in the following sections, the full \acp{HRTF} needed to be reconstructed, including the additional phase information that was disregarded in the pre-processing step. This phase information was removed on the assumption that the up-sampled \acp{HRTF} can be reconstructed using a minimum-phase approximation and a simple ITD model.

This minimum-phase approximation is achieved by calculating the minimum-phase function, $m(\omega)$, which can be uniquely determined by the magnitude spectrum of each transfer function, $H(\omega)$, at every source position in the \ac{HRTF} through the Hilbert transform \cite{Mehrgardt1977}
\begin{equation}
m(\omega)=\mathscr{H}\{-\ln(|H(\omega)|)\}\;,
\end{equation}
where $\mathscr{H}\{\cdot\}$ denotes the Hilbert transform, and $\omega$ represents the frequency bin.

The simple model used to calculate the \acp{ITD} is based on the radius, $r$, of the listener's head (set to  8.75~cm, which corresponds to an average adult head size), the speed of sound, $c$, (approximately 343~m/s) and the interaural azimuth, $\theta_I$, (in radians, from 0 to $\frac{\pi}{2}$ for sources on the listener's left, and from $\frac{\pi}{2}$ to $\pi$ for sources on the listener's right) using
\begin{equation}
 \begin{aligned}
    \text{ITD} &= \frac{r}{c}\big(\theta_I + \sin(\theta_I)\big)\;, \\ \text{where} \quad \theta_I &= \arcsin\big(\sin(\theta)\cos(\phi)\big)\;.
 \end{aligned}
 \label{eq:itd}
\end{equation}

It should be noted that a final interpolation could also be performed to map the points generated from the gnomonic equiangular projection back to an even spherical distribution. The main reason why this final interpolation is not performed is because it is unnecessary. As sofa files are all measured on different grids, most, if not all, software that utilises sofa files re-interpolates them onto a uniform grid \cite{Cuevas-Rodriguez2019} before deployment. In the proposed method, the grid of the upsampled \acp{HRTF} is very dense. Therefore, any re-interpolation that may take place in spatial acoustic software would only introduce errors that would not be perceived perceptually.
\begin{figure}[t]
\centering

  \subfloat[Original.\label{fig:sphere_LR_1280}]{
  \includegraphics[clip, trim=1.7cm 1.6cm 1.7cm 1.9cm, width=0.25\linewidth,keepaspectratio]{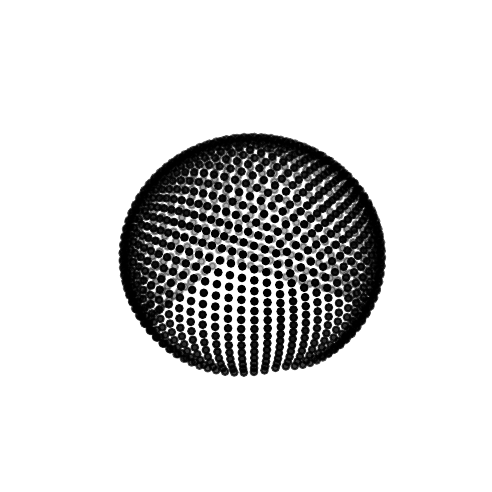}
} \hspace{0.2cm} 
  \subfloat[1280 Target.\label{fig:cube_LR_1280}]{
  \includegraphics[clip, trim=0.8cm 0.8cm 0.8cm 1.2cm, width=0.25\linewidth,keepaspectratio]{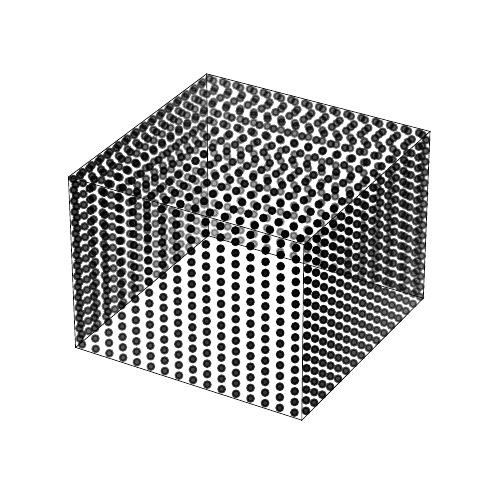}
} \hspace{0.2cm} 
  \subfloat[320 $\,\rightarrow$ 1280.\label{fig:cube_LR_320}]{
  \includegraphics[clip, trim=0.8cm 0.8cm 0.8cm 1.2cm, width=0.25\linewidth,keepaspectratio]{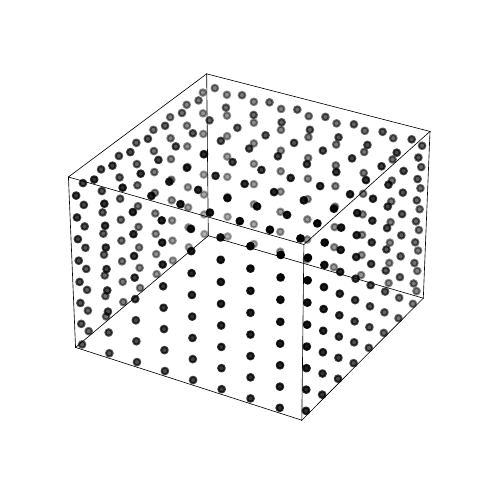}
} 

  \subfloat[80 $\,\rightarrow$ 1280.\label{fig:cube_LR_80}]{%
  \includegraphics[clip, trim=0.8cm 0.8cm 0.8cm 1.2cm, width=0.25\linewidth,keepaspectratio]{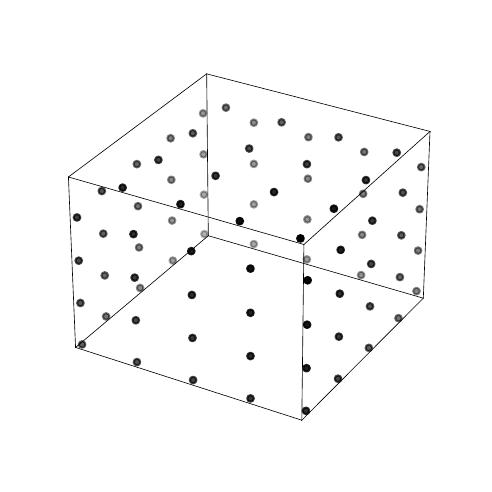}
} \hspace{0.2cm}
  \subfloat[20 $\,\rightarrow$ 1280.\label{fig:cube_LR_20}]{
  \includegraphics[clip, trim=0.8cm 0.8cm 0.8cm 1.2cm, width=0.25\linewidth,keepaspectratio]{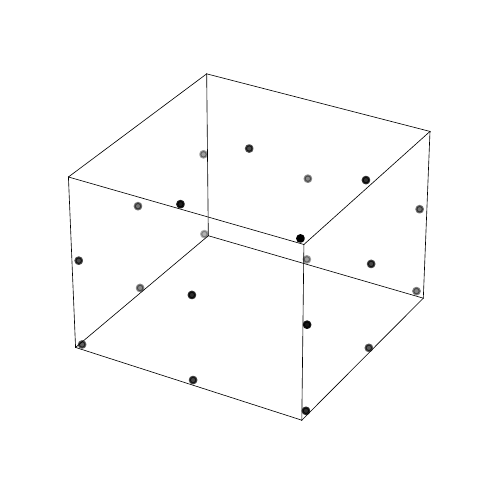}
} \hspace{0.2cm}
  \subfloat[5 $\,\rightarrow$ 1280.\label{fig:cube_LR_5}]{%
  \includegraphics[clip, trim=0.8cm 0.8cm 0.8cm 1.2cm, width=0.25\linewidth,keepaspectratio]{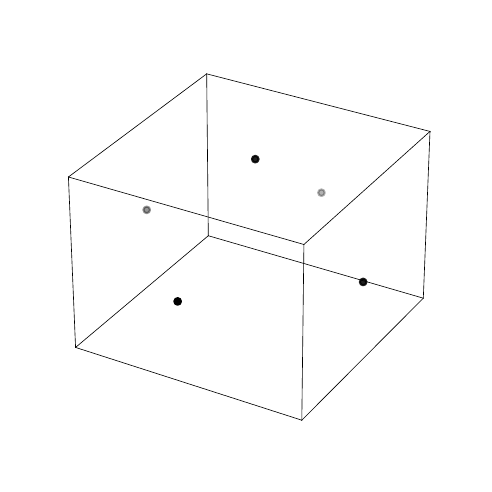}
}
 \caption{The source positions for each downsampling factor.}
\label{fig:sphere_LR_positions}
\end{figure}

\section{Experimental Setup}
\label{sec:experimental_setup}
\subsection{Data}

The network was trained and validated on the \ac{HRTF} dataset measured in Austria by the \ac{ARI} and, throughout this paper, will be referred to as the \ac{ARI} \ac{HRTF} database \cite{Majdak2022}. The dataset contains \ac{HRTF} measurements on 221 subjects for both the left and right ear (442 \ac{HRTF} in total for both ears), making it one of the largest measured \ac{HRTF} datasets available. Each measured \ac{HRTF} contains \acp{IR} for 1550 directions around the listener; these range from 0$^\circ$ to 360$^\circ$ in terms of azimuth and from -30$^\circ$ to 80$^\circ$ in terms of elevation (a measurement for the top position was not available in this specific dataset). The number of measurements near the horizontal plane was of a higher density, which is common in \ac{HRTF} measurement systems to reflect that humans can localise sounds in this space more accurately. The \ac{ARI} \ac{HRTF} dataset is read using the \code{Hartufo} toolkit \cite{Pauwels2023, AES2022}, which was developed for \ac{HRTF} data management with a specific focus on deep learning.

After processing all \acp{HRTF} in the \ac{ARI} \ac{HRTF} dataset as described in \cref{sec:data_pre_processing}, the 442 \acp{HRTF} are split so that 352 \acp{HRTF} are used for training and 90 \acp{HRTF} are used for validation. This represents an 80-20 split between training and validation sets where the left and right ear for the same individual are not split between sets, i.e. a subject may not contribute one ear to training and another ear to the validation. This ensures that the generator network is tested on unseen data from a given individual.

\subsection{Training}

The high-resolution 1280 target is generated by pre-processing the ARI \ac{HRTF} dataset (which contains 1550 positions for each listener) into 5 panels, which contain 16 by 16 source positions (i.e. 1280 positions~$=5$~$\times$~16~$\times$~16, shown in \crefsub{fig:projected_data_flattened_interpolation}) where the Kalman filter parameters where set to $\eta=\num{5e-3}$, $\sigma^2_w=\num{1e-4}$, $\sigma^2_v=400$ and the \acp{IR} where trimmed to 10 samples before the onset and to a length of 128 samples.

To obtain the low-resolution \ac{HRTF} inputs from the high-resolution targets, the \acp{HRTF} are downsampled by selecting one \ac{IR} in every $r$. This means each high-resolution \ac{HRTF} target, generated from the gnomonic equiangular projection, whose dimensions are $256\times5\times16\times16$, are downsampled to $256\times5\times\frac{16}{r}\times\frac{16}{r}$ to create each input. In the case where only 5 source positions are available, the centre position of each panel is selected with coordinates (8,8).  Therefore, the generator network aims to preserve the points of the low-resolution \ac{HRTF} given at the input while interpolating all the other points to match the target. 

It should be noted that the $256$ dimension refers to the concatenation of the two left and right ear 128-point \acp{IR}. \cref{fig:sphere_LR_positions} shows the positions of the sources for $r$ values 2 (320~$\,\rightarrow$~1280), 4 (80~$\,\rightarrow$~1280), 8 (20~$\,\rightarrow$~1280), 16 (5~$\,\rightarrow$~1280). These positions were selected using the \code{torch.nn.functional.interpolate} function, where the `\code{scale\_factor}' was set to $r$. The high-resolution \ac{HRTF} target of 1280 positions (i.e. $5\times16\times16$) was selected as it is comparable to the 1550 positions measured in the \ac{ARI} \ac{HRTF} dataset. 

It should also be noted that the 128-point \ac{FFT} magnitude inputs are also not scaled or normalised, as the \ac{LSD} metric used in the content loss for the generator, $G$, requires non-normalised magnitudes. 

The \acp{GAN} hyperparameters were adjusted in order to find the best-performing model in terms of the cost function on the training data. To achieve this hyperparameter tuning, a grid search was deployed using \cite{Liaw2018}, where the search space consisted of `{\Ac{LR}} - Generator': \{\num{2.0e-4}, \num{4.0e-4}, \num{6.0e-4}, \num{8.0e-4}\},
`\Ac{LR} - Discriminator': \{\num{1.5e-6}, \num{3.0e-6}, \num{4.5e-6}, \num{6.0e-6}\},
`Number of epochs': \{300, 250, 200, 150\},
`Adversarial weight ($\lambda_\text{A}$)': \{0.1, 0.01, 0.001\},
`Content weight ($\lambda_\text{C}$)': \{0.1, 0.01, 0.001\}. This resulted in a variation of 15.5\% in terms of the cost function with the selected values for each of the four networks given in \cref{tab:hyperparameters}.
% A sweep of hyperparameters (learning rates and loss component weights) was performed, but only a 15.5\% variation in the performance was measured. 
Various batch sizes were also explored informally; a relatively small batch size of 8 was found to give the best performance.

\begin{table}[t]
\centering
\renewcommand{\arraystretch}{1.19}
\setlength{\tabcolsep}{3.0pt}
\caption{The hyperparameter values selected for the four networks with different upsampling factors. \label{tab:hyperparameters}}
\resizebox{0.95\linewidth}{!}{%
\begin{tabular}{|c|c @{\hspace{-0.3\tabcolsep}}|c|c|c|c|}
\hhline{-~----}
\multirow{2}{*}{\textbf{Hyperparameter}} & & \multicolumn{4}{c|}{\textbf{Upsample Factor [No. orginal  $\,\rightarrow$ No. upsampled]}}                                                               \\ \cline{3-6}
                        & & \multicolumn{1}{c|}{ \textbf{320} $\,\rightarrow$ \textbf{1280}} & \multicolumn{1}{c|}{ \textbf{80} $\,\rightarrow$ \textbf{1280}} & \multicolumn{1}{c|}{ \textbf{20} $\,\rightarrow$ \textbf{1280}} & \multicolumn{1}{c|}{ \textbf{5} $\,\rightarrow$ \textbf{1280}} \\ \hhline{=~====}
\textbf{No. Epochs}             & & 300 & 300 & 300 & 300   \\ \hhline{-~----}
\textbf{LR\,-\,Generator}             & & \num{2.0e-4} & \num{8.0e-4} & \num{2.0e-4} & \num{2.0e-4}   \\ \hhline{-~----}
\textbf{LR\,-\,Discriminator}             & & \num{1.5e-06} & \num{1.5e-06} & \num{1.5e-06} & \num{1.5e-06}   \\ \hhline{-~----}
\textbf{Content Weight} ($\lambda_\text{C}$)             & & 0.1 & 0.01 & 0.001 & 0.01   \\ \hhline{-~----}
\textbf{Adversarial Weight} ($\lambda_\text{A}$)             & & 0.001 & 0.1 & 0.001 & 0.01   \\ \hhline{-~----}
\end{tabular}
}
\end{table}

The model was trained using the \textit{Adam} optimiser \cite{Kingma2015}, with hyperparameter values of $\beta_1 = 0.9$ and $\beta_2 = 0.999$. In training, $D$ and $G$ were alternately updated with different frequencies to improve stability, with $D$ being updated four times for every $G$ update. This was found by informally testing different ratios inspired by \cite{Gulrajani2017}. The kernel weights were initialised using Kaiming initialisation \cite{He2015a}. This model was implemented using a PyTorch framework with custom modules for gnomonic equiangular projection padding and convolution layers and trained on an NVIDIA Quadro RTX 6000 \ac{GPU}.

\subsection{Convergence}
In \cref{fig:Overall_loss_curves}, the convergence of both the generator and the discriminator networks is given for the 20 $\,\rightarrow$ 1280 network. It should be noted that owing to the fact that a \ac{GAN} consists of two networks competing against each other, an improvement in the generator will lead to a higher loss in the discriminator and vice versa. Hence, both networks converge to a stable value over time (although this is not expected to be zero). It can also be seen in \cref{fig:Overall_loss_curves} that the generator network can converge quicker than the discriminator. Although not shown here, similar loss curves can be observed for the three other networks (5~$\,\rightarrow$~1280, 80~$\,\rightarrow$~1280 and 320~$\,\rightarrow$~1280).

\textbf{\begin{figure}[tb]
\centering
\includegraphics[width=\linewidth]{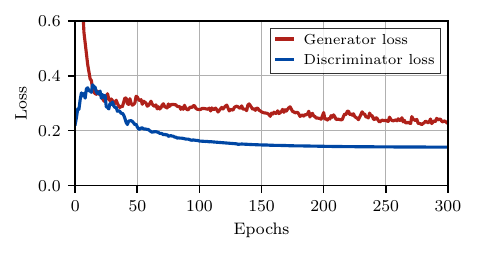}
% \vspace{-0.7cm}
 \caption{Illustrative example of overall loss curves for 20 $\,\rightarrow$ 1280 network.}
\label{fig:Overall_loss_curves}
\end{figure}}

\subsection{Baselines}
Three baseline methods have been used as a benchmark for the comparison of the results from the experimental evaluation.

\subsubsection{Baseline-1 - SH interpolation}
An approach that has shown to yield good performance in the domain of \ac{HRTF} upsampling is \ac{SH} interpolation. It works by projecting the \ac{HRTF} onto a set of spherical basis functions, known as spherical harmonics, which produces a continuous representation of an \ac{HRTF} \cite{Arend2021a}. This work uses a magnitude-corrected and time-aligned \ac{SH} interpolation presented in \cite{Arend2023b} as a baseline. 

\subsubsection{Baseline-2 - Barycentric interpolation}
The most common method for upsampling is that of barycentric interpolation, which is used in this section as a baseline and described in \cref{sec:barycentric_interpolation}. It should be noted, however, that the approach in \cref{sec:barycentric_interpolation} is modified when the number of source positions is only 5 in the low-resolution \ac{HRTF}. This is because if no triangle can be formed around the point to be interpolated, which is sometimes the case at this low resolution, then the three closest points need to be used instead.

\subsubsection{Baseline-3 - non-individual HRTF selection} As mentioned in \cref{sec:intro}, instead of interpolating a low-resolution \ac{HRTF}, we can just select an \ac{HRTF} from a database. In this baseline, instead of randomly selecting an \ac{HRTF}, two \acp{HRTF} are selected from the training dataset based on their average \ac{LSD} error when comparing them against all other \acp{HRTF} in the training set. The subject whose \ac{HRTF} produces the lowest average \ac{LSD} error is considered the most generic (Selection-1), and the subject whose \ac{HRTF} produces the largest average \ac{LSD} error is considered the most unique (Selection-2). It must be highlighted, however, that this selection is only based on the \ac{LSD}, and not all \ac{LSD} errors have the same perceptual relevance.

\section{Experimental evaluation}
\label{sec:results}

This section will compare four different levels of \ac{SRGAN} upsampling against the two baselines described earlier for 45 test subjects. These upsampling levels include 320, 80, 20 and 5 source positions to 1280 source positions. 

% To implement the layers of the \ac{SRGAN}, custom PyTorch modules were developed for this work, based on the custom Keras \cite{Chollet2015} modules introduced in \cite{Weyn2020}. 
The complete \ac{SRGAN} implementation and pre-processing code to reproduce these results are available at \cite{Hogg2023}.

\begin{table}[tb]
\centering
\renewcommand{\arraystretch}{1.19}
\setlength{\tabcolsep}{3.0pt}
\caption{A comparison of the mean \acf{LSD} and (\acf{SD}) error across all source positions for different upsampling factors. The `best' performance of each upsampling factor has been highlighted. \label{tab:sd_error}}
\resizebox{0.95\linewidth}{!}{%
% \begin{tabular}{|c|c @{\hspace{-0.3\tabcolsep}}|c|c|c|c|}
% \hhline{-~----}
% \multirow{2}{*}{\textbf{Method}} & & \multicolumn{4}{c|}{\textbf{Upsample Factor [No. orginal  $\,\rightarrow$ No. upsampled]}}                                                               \\ \cline{3-6}
%                         & & \multicolumn{1}{c|}{ \textbf{320} $\,\rightarrow$ \textbf{1280}} & \multicolumn{1}{c|}{ \textbf{80} $\,\rightarrow$ \textbf{1280}} & \multicolumn{1}{c|}{ \textbf{20} $\,\rightarrow$ \textbf{1280}} & \multicolumn{1}{c|}{ \textbf{5} $\,\rightarrow$ \textbf{1280}} \\ \hhline{=~====}
% \textbf{SRGAN}             & & 3.13 (0.14) & 4.32 (0.25)  & 4.91 (0.27)  & 5.25 (0.35)   \\ \hhline{-~----}
% \textbf{Barycentric}       & & 2.40 (0.20) & 3.62 (0.22)  & 5.16 (0.23)  & 6.91 (0.37)   \\ \hhline{=~====}
% \textbf{Selection-1}             & & \multicolumn{4}{c|}{6.90 (0.49)}   \\ \hhline{-~----}
% \textbf{Selection-2}             & & \multicolumn{4}{c|}{8.25 (0.49)}   \\ \hhline{-~----}
% \end{tabular}
\begin{tabular}{|c|c @{\hspace{-0.3\tabcolsep}}|c|c|c|c|}
\hhline{-~----}
\multirow{2}{*}{\textbf{Method}} & & \multicolumn{4}{c|}{\textbf{Upsampling [No. orginal  $\,\rightarrow$ No. upsampled]}}                                                               \\ \cline{3-6}
                        & & \multicolumn{1}{c|}{ \textbf{320} $\,\rightarrow$ \textbf{1280}} & \multicolumn{1}{c|}{ \textbf{80} $\,\rightarrow$ \textbf{1280}} & \multicolumn{1}{c|}{ \textbf{20} $\,\rightarrow$ \textbf{1280}} & \multicolumn{1}{c|}{ \textbf{5} $\,\rightarrow$ \textbf{1280}} \\ \hhline{=~====}
% \textbf{SRGAN}             & & 3.13 (0.14) & 4.32 (0.25)  & \cellcolor{green!20} 4.91 (0.27)  & \cellcolor{green!20} 5.29 (0.34)   \\ \hhline{-~----}
% \textbf{Barycentric}       & & \cellcolor{green!20} 2.40 (0.20) & \cellcolor{green!20} 3.62 (0.22)  & 5.16 (0.23)  & 7.96 (0.37)   \\ \hhline{=~====}
% \textbf{Selection-1}             & & \multicolumn{4}{c|}{6.90 (0.49)}   \\ \hhline{-~----}
% \textbf{Selection-2}             & & \multicolumn{4}{c|}{8.25 (0.49)}   \\ \hhline{-~----}

\textbf{SRGAN} & & 3.28 (0.13) & 4.86 (0.24) & \cellcolor{green!20} 4.99 (0.27) & \cellcolor{green!20} 5.30 (0.35) \\ \hhline{=~====}
\textbf{SH} & & 3.54 (0.15) & 4.94 (0.20) & 5.90 (0.25) & 10.36 (0.74) \\ \hhline{-~----}
\textbf{Barycentric} & & \cellcolor{green!20} 2.50 (0.20) & \cellcolor{green!20} 3.71 (0.22) & 5.18 (0.23) & 7.30 (0.33) \\ \hhline{=~====}
\textbf{Selection-1} & & \multicolumn{4}{c|}{6.96 (0.47)} \\ \hhline{-~----}
\textbf{Selection-2} & & \multicolumn{4}{c|}{8.20 (0.61)} \\ \hhline{-~----}

\end{tabular}
}
\end{table}
% \begin{table}[tb]
% \centering
% \renewcommand{\arraystretch}{1.19}
% \setlength{\tabcolsep}{3.0pt}
% \caption{A comparison of the mean \acf{LSD} and (\acf{SD}) on the original positions after SRGAN post-processing. The mean error is across all source positions for different upsampling factors. The `best' performance of each upsampling factor has been highlighted. \label{tab:sd_error_post_processing}}
% \resizebox{0.95\linewidth}{!}{%
% \begin{tabular}{|c|c @{\hspace{-0.3\tabcolsep}}|c|c|c|c|}
% \hhline{-~----}
% \multirow{2}{*}{\textbf{Method}} & & \multicolumn{4}{c|}{\textbf{Upsample Factor (No. original  $\,\rightarrow$ upsampled) [dB]}} \\ \hhline{~~----}
% & & \multicolumn{1}{c|}{ \textbf{320} $\,\rightarrow$ \textbf{1280}} & \multicolumn{1}{c|}{ \textbf{80} $\,\rightarrow$ \textbf{1280}} & \multicolumn{1}{c|}{ \textbf{20} $\,\rightarrow$ \textbf{1280}} & \multicolumn{1}{c|}{ \textbf{5} $\,\rightarrow$ \textbf{1280}} \\ \hhline{=~====}
% \textbf{SRGAN} & & 3.75 (0.17) & 4.63 (0.24) & 4.70 (0.28) & \cellcolor{green!20} 4.79 (0.28) \\ \hhline{-~----}
% \textbf{SH} & & \cellcolor{green!20} 2.61 (0.16) & 3.74 (0.20) & 6.60 (0.31) & 11.80 (0.41) \\ \hhline{-~----}
% \textbf{Barycentric} & & 3.30 (0.21) & \cellcolor{green!20} 3.64 (0.20) & \cellcolor{green!20} 4.52 (0.21) & 5.97 (0.32) \\ \hhline{-~----}
% \textbf{Selection-1} & & \multicolumn{4}{c|}{5.73 (0.47)} \\ \hhline{-~----}
% \textbf{Selection-2} & & \multicolumn{4}{c|}{6.94 (0.66)} \\ \hhline{-~----}
% \end{tabular}}
% \end{table}
\begin{figure}[tb]
\centering
\includegraphics[width=\linewidth]{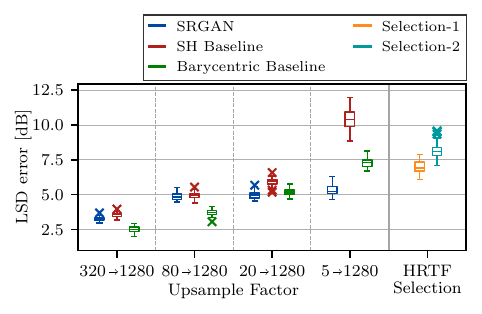}
\vspace{-0.7cm}
 \caption{\Acf{LSD} error comparison.}
\label{fig:sd_error}
\end{figure}

\newcommand{\heightFig}{4.93cm}
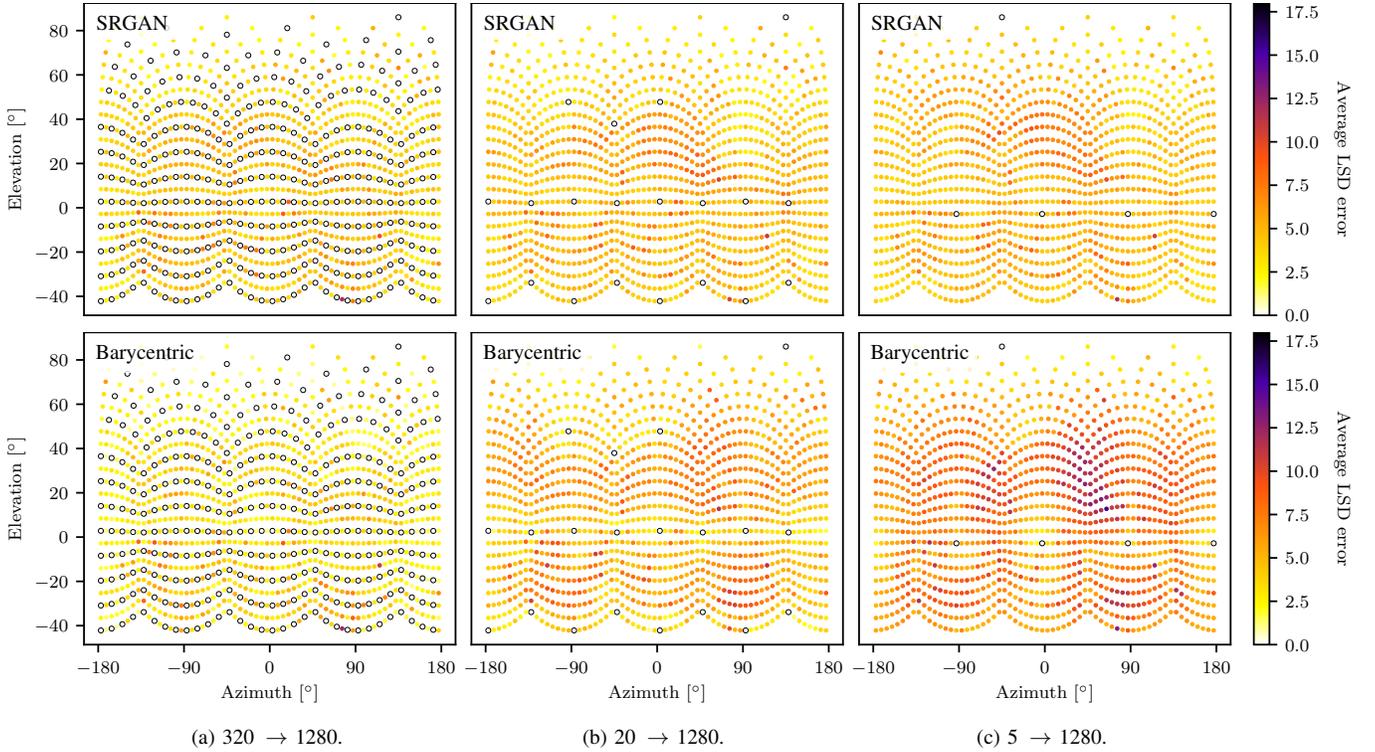
\begin{figure*}[tb]
\captionsetup[subfloat]{captionskip=9pt} 
  \centering
  \subfloat{
   \begin{tikzpicture}
   \node[inner sep=0pt] (img) {\includegraphics[clip, trim=0cm 0cm 1cm 0cm, height=\heightFig,keepaspectratio]{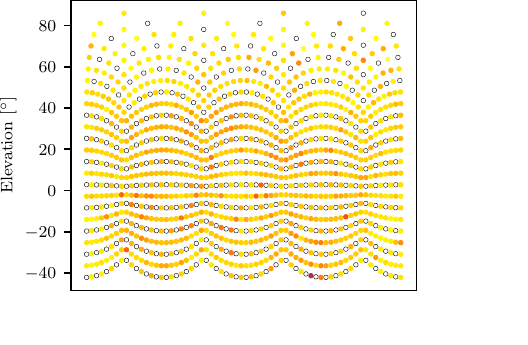}};
   \node [below left,align=center, fill=white, fill opacity=0.8, text opacity= 1.0] at ([yshift=-0.05cm, xshift=-4.45cm]img.north east){\footnotesize{SRGAN}};
  \end{tikzpicture}
} \hspace{-0.95cm}
  % \subfloat[Discriminator loss curves.\label{fig:D_loss_curves}]{%
  % \includegraphics[height=\heightFig,keepaspectratio]{figures/868_SD_srgan_interpolated_80_1280.pdf}
% }
  \subfloat{%
    \begin{tikzpicture}
   \node[inner sep=0pt] (img) {\includegraphics[clip, trim=1cm 0cm 1cm 0cm, height=\heightFig,keepaspectratio]{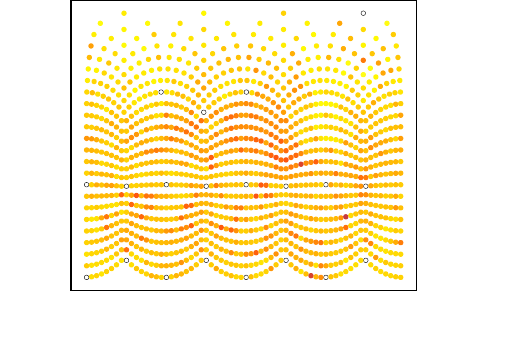}};
   \node [below left,align=center, fill=white, fill opacity=0.8, text opacity= 1.0] at ([yshift=-0.05cm, xshift=-4.45cm]img.north east){\footnotesize{SRGAN}};
  \end{tikzpicture}
} \hspace{-1.07cm}
  \subfloat{
     \begin{tikzpicture}
   \node[inner sep=0pt] (img) {\includegraphics[clip, trim=1cm 0cm 0cm 0cm, height=\heightFig,keepaspectratio]{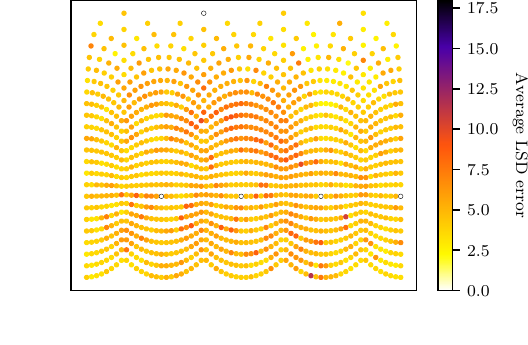}};
   \node [below left,align=center, fill=white, fill opacity=0.8, text opacity= 1.0] at ([yshift=-0.05cm, xshift=-5.3cm]img.north east){\footnotesize{SRGAN}};
  \end{tikzpicture}
} 
\vspace{-0.9cm}\setcounter{subfigure}{0}
\captionsetup[subfigure]{oneside,margin={0.2cm,0cm}}
  \subfloat[320 $\,\rightarrow$ 1280.\label{fig:sd_error_all_nodes_320}]{
  \begin{tikzpicture}
   \node[inner sep=0pt] (img) {\includegraphics[clip, trim=0cm 0cm 1cm 0cm, height=\heightFig,keepaspectratio]{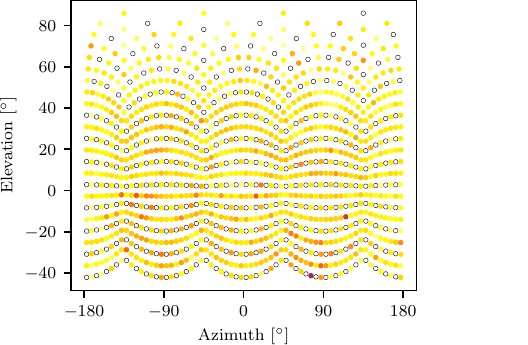}};
   \node [below left,align=center, fill=white, fill opacity=0.8, text opacity= 1.0] at ([yshift=-0.05cm, xshift=-4.1cm]img.north east){\footnotesize{Barycentric}};
  \end{tikzpicture}
} \hspace{-0.95cm}
  % \subfloat[Discriminator loss curves.\label{fig:D_loss_curves}]{%
  % \includegraphics[height=\heightFig,keepaspectratio]{figures/868_SD_barycentric_interpolated_80_1280.pdf}
% }
\captionsetup[subfigure]{oneside,margin={-0.8cm,0cm}}
  \subfloat[20 $\,\rightarrow$ 1280.\label{fig:sd_error_all_nodes_20}]{%
  \begin{tikzpicture}
   \node[inner sep=0pt] (img) {\includegraphics[clip, trim=1cm 0cm 1cm 0cm, height=\heightFig,keepaspectratio]{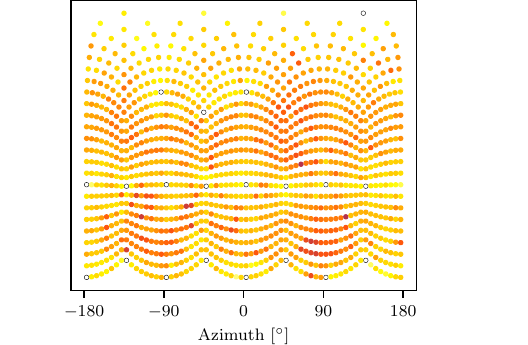}};
   \node [below left,align=center, fill=white, fill opacity=0.8, text opacity= 1.0] at ([yshift=-0.05cm, xshift=-4.1cm]img.north east){\footnotesize{Barycentric}};
  \end{tikzpicture}
} \hspace{-1.07cm}
  \captionsetup[subfigure]{oneside,margin={-1.5cm,0cm}}
  \subfloat[5 $\,\rightarrow$ 1280.\label{fig:sd_error_all_nodes_5}]{
   \begin{tikzpicture}
   \node[inner sep=0pt] (img) {\includegraphics[clip, trim=1cm 0cm 0cm 0cm, height=\heightFig,keepaspectratio]{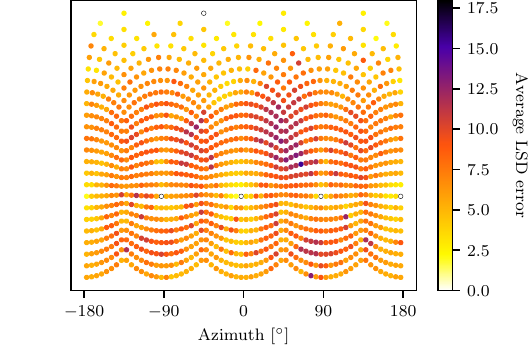}};
   \node [below left,align=center, fill=white, fill opacity=0.8, text opacity= 1.0] at ([yshift=-0.05cm, xshift=-4.95cm]img.north east){\footnotesize{Barycentric}};
  \end{tikzpicture}

} 
\medskip
\caption{Comparison of the proposed \acf{SRGAN} (top) and barycentric (bottom) in terms of \acf{LSD} errors at different levels of upsampling SubjectID 868 (all source positions). The original source positions before interpolation are outlined with a black circle.}
\label{fig:sd_error_all_nodes}
\end{figure*}
\begin{figure*}[tb]
\captionsetup[subfloat]{captionskip=9pt} 
\newcommand{\heightFigTwo}{4.93cm}
  \centering
  \captionsetup[subfigure]{oneside,margin={0.1cm,0cm}}
  \subfloat[The most `generic' \ac{HRTF}.\label{fig:sd_error_all_nodes_min}]{
   \begin{tikzpicture}
   \node[inner sep=0pt] (img) {\includegraphics[clip, trim=0cm 0cm 1cm 0cm, height=\heightFigTwo,keepaspectratio]{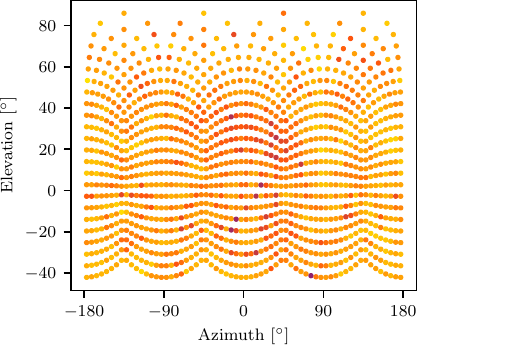}};
   \node [below left,align=center, fill=white, fill opacity=0.8, text opacity= 1.0] at ([yshift=-0.05cm, xshift=-4.13cm]img.north east){\footnotesize{Selection-1}};
  \end{tikzpicture}
} \hspace{-0.85cm}
  \captionsetup[subfigure]{oneside,margin={-1.8cm,0cm}}
  \subfloat[The most `unique' \ac{HRTF}.\label{fig:sd_error_all_nodes_max}]{%
   \begin{tikzpicture}
   \node[inner sep=0pt] (img) {\includegraphics[clip, trim=1.1cm 0cm 0cm 0cm, height=\heightFigTwo,keepaspectratio]{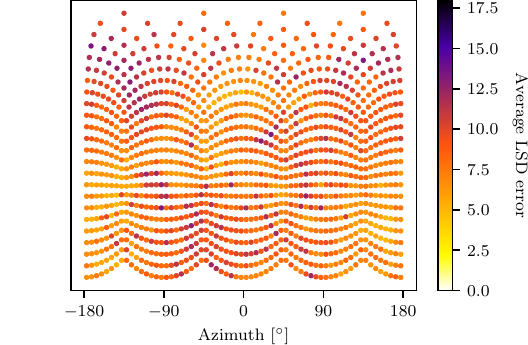}};
   \node [below left,align=center, fill=white, fill opacity=0.8, text opacity= 1.0] at ([yshift=-0.05cm, xshift=-4.97cm]img.north east){\footnotesize{Selection-2}};
  \end{tikzpicture}
} 
\medskip
\caption{The \acf{LSD} of Basline-2 showing the comparison between Selection-1 and Selection-2 for SubjectID 868 (all source positions).}
\label{fig:sd_error_all_nodes_min_max}
\end{figure*}

\begin{table}[!tb]
\centering
\renewcommand{\arraystretch}{1.19}
\captionsetup[subfloat]{captionskip=10pt} 
\setlength{\tabcolsep}{3.0pt}
\caption{The mean and (\acf{SD}) values of the model-based perceptual evaluation across the subjects in the test set for the different upsampling factors. The `best' performance of each upsampling factor has been highlighted. \label{tab:perceptual_evaluation}}
\subfloat[Polar accuracy error comparison. \label{tab:p_acc_error}]{
\resizebox{0.95\linewidth}{!}{%
\begin{tabular}{|c|c @{\hspace{-0.3\tabcolsep}}|c|c|c|c|}
\hhline{-~----}
\multirow{2}{*}{\textbf{Method}} & & \multicolumn{4}{c|}{\textbf{Upsampling [No. orginal  $\,\rightarrow$ No. upsampled]}}                                                               \\ \cline{3-6}
                        & & \multicolumn{1}{c|}{ \textbf{320} $\,\rightarrow$ \textbf{1280}} & \multicolumn{1}{c|}{ \textbf{80} $\,\rightarrow$ \textbf{1280}} & \multicolumn{1}{c|}{ \textbf{20} $\,\rightarrow$ \textbf{1280}} & \multicolumn{1}{c|}{ \textbf{5} $\,\rightarrow$ \textbf{1280}} \\ \hhline{=~====}
% \textbf{SRGAN}             & & 1.60 (4.92) & -3.61 (7.48)  & \cellcolor{green!20} 0.74 (8.20)  & \cellcolor{green!20} 0.24 (13.81)   \\ \hhline{-~----}
% \textbf{Barycentric}       & &  \cellcolor{green!20} 0.72 (4.29) & \cellcolor{green!20} 0.93 (4.84)  & 0.36 (8.74)  & -15.03 (41.78)   \\ \hhline{=~====}
% \textbf{Selection-1}             & & \multicolumn{4}{c|}{0.21 (13.90)}   \\ \hhline{-~----}
% \textbf{Selection-2}             & & \multicolumn{4}{c|}{-12.82 (16.52)}   \\ \hhline{=~====}
% \textbf{Target}             & & \multicolumn{4}{c|}{0.76 (4.07)}   \\ \hhline{-~----}

\textbf{SRGAN} & & 0.46 (4.66) & 2.73 (6.82) & 1.06 (10.02) & \cellcolor{green!20} -0.70 (8.33) \\ \hhline{=~====}
\textbf{SH} & & \cellcolor{green!20} -0.39 (4.38) & -1.92 (4.54) & -4.17 (5.02) & -28.61 (17.55) \\ \hhline{-~----}
\textbf{Barycentric} & & 1.17 (3.84) & \cellcolor{green!20} 1.57 (4.32) & \cellcolor{green!20} 2.22 (8.36) & -2.54 (23.84) \\ \hhline{=~====}
\textbf{Selection-1} & & \multicolumn{4}{c|}{2.05 (17.17)} \\ \hhline{-~----}
\textbf{Selection-2} & & \multicolumn{4}{c|}{22.18 (21.75)} \\ \hhline{=~====}
\textbf{Target} & & \multicolumn{4}{c|}{0.86 (3.74)} \\ \hhline{-~----}
\end{tabular}
}}\medskip

\subfloat[Quadrant error comparison. \label{tab:quadrant_error}]{
\resizebox{0.95\linewidth}{!}{%
\begin{tabular}{|c|c @{\hspace{-0.3\tabcolsep}}|c|c|c|c|}
\hhline{-~----}
\multirow{2}{*}{\textbf{Method}} & & \multicolumn{4}{c|}{\textbf{Upsampling [No. orginal  $\,\rightarrow$ No. upsampled]}}                                                               \\ \cline{3-6}
                        & & \multicolumn{1}{c|}{ \textbf{320} $\,\rightarrow$ \textbf{1280}} & \multicolumn{1}{c|}{ \textbf{80} $\,\rightarrow$ \textbf{1280}} & \multicolumn{1}{c|}{ \textbf{20} $\,\rightarrow$ \textbf{1280}} & \multicolumn{1}{c|}{ \textbf{5} $\,\rightarrow$ \textbf{1280}} \\ \hhline{=~====}
% \textbf{SRGAN}             & & 9.68 (3.09) & 12.62 (3.85)  & \cellcolor{green!20} 13.92 (4.54)  &  20.87 (9.24)   \\ \hhline{-~----}
% \textbf{Barycentric}       & & \cellcolor{green!20} 8.96 (2.86) & \cellcolor{green!20} 9.59 (2.92)  & 14.87 (3.84)  & 37.70 (5.20)   \\ \hhline{=~====}
% \textbf{Selection-1}             & & \multicolumn{1}{c}{} & \multicolumn{2}{c}{  20.22 (8.96)} & \cellcolor{green!20}   \\ \hhline{-~----}
% \textbf{Selection-2}             & & \multicolumn{4}{c|}{21.72 (10.40)}   \\ \hhline{=~====}
% \textbf{Target}             & & \multicolumn{4}{c|}{8.74 (2.74)}   \\ \hhline{-~----}

\textbf{SRGAN} & & 8.64 (2.78) & 9.83 (3.25) & 16.53 (4.59) & \cellcolor{green!20} 11.28 (3.60) \\ \hhline{=~====}
\textbf{SH} & & \cellcolor{green!20} 8.43 (2.89) & 10.10 (2.99) & 16.04 (3.66) & 19.17 (8.25) \\ \hhline{-~----}
\textbf{Barycentric} & & 8.50 (2.68) & \cellcolor{green!20} 9.15 (2.73) & \cellcolor{green!20} 13.79 (3.76) & 24.65 (7.28) \\ \hhline{=~====}
\textbf{Selection-1} & & \multicolumn{4}{c|}{22.36 (7.59)} \\ \hhline{-~----}
\textbf{Selection-2} & & \multicolumn{4}{c|}{22.17 (12.14)} \\ \hhline{=~====}
\textbf{Target} & & \multicolumn{4}{c|}{7.99 (2.76)} \\ \hhline{-~----}
\end{tabular}
}}\medskip

\subfloat[Polar \acf{RMS} error comparison. \label{tab:p_rms_error}]{
\resizebox{0.95\linewidth}{!}{%
\begin{tabular}{|c|c @{\hspace{-0.3\tabcolsep}}|c|c|c|c|}
\hhline{-~----}
\multirow{2}{*}{\textbf{Method}} & & \multicolumn{4}{c|}{\textbf{Upsampling [No. orginal  $\,\rightarrow$ No. upsampled]}}                                                               \\ \cline{3-6}
                        & & \multicolumn{1}{c|}{ \textbf{320} $\,\rightarrow$ \textbf{1280}} & \multicolumn{1}{c|}{ \textbf{80} $\,\rightarrow$ \textbf{1280}} & \multicolumn{1}{c|}{ \textbf{20} $\,\rightarrow$ \textbf{1280}} & \multicolumn{1}{c|}{ \textbf{5} $\,\rightarrow$ \textbf{1280}} \\ \hhline{=~====}
% \textbf{SRGAN}             & & \cellcolor{green!20} 33.04 (1.97) & 34.94 (2.15)  & \cellcolor{green!20} 36.34 (1.78)  & \cellcolor{green!20} 37.21 (3.04)   \\ \hhline{-~----}
% \textbf{Barycentric}       & & 33.13 (1.80) &  \cellcolor{green!20} 34.35 (1.83)  & 39.22 (1.24)  & 46.16 (1.06)   \\ \hhline{=~====}
% \textbf{Selection-1}             & & \multicolumn{4}{c|}{38.30 (2.97)}   \\ \hhline{-~----}
% \textbf{Selection-2}             & & \multicolumn{4}{c|}{39.80 (3.45)}   \\ \hhline{=~====}
% \textbf{Target}             & & \multicolumn{4}{c|}{32.88 (1.82)}   \\ \hhline{-~----}

\textbf{SRGAN} & & 32.96 (1.83) & 35.46 (1.73) & 36.50 (1.65) & \cellcolor{green!20} 36.03 (1.87) \\ \hhline{=~====}
\textbf{SH} & & 33.02 (2.04) & 33.83 (2.21) & \cellcolor{green!20} 34.71 (2.24) & 41.98 (2.52) \\ \hhline{-~----}
\textbf{Barycentric} & & \cellcolor{green!20} 32.61 (1.70) & \cellcolor{green!20} 33.75 (1.68) & 38.24 (1.36) & 41.79 (1.22) \\ \hhline{=~====}
\textbf{Selection-1} & & \multicolumn{4}{c|}{38.89 (2.43)} \\ \hhline{-~----}
\textbf{Selection-2} & & \multicolumn{4}{c|}{39.90 (2.32)} \\ \hhline{=~====}
\textbf{Target} & & \multicolumn{4}{c|}{32.11 (2.10)} \\ \hhline{-~----}
\end{tabular}
}}
\end{table}

\newcommand{\widthFig}{0.94\linewidth}
\newcommand\trimTop{1.3}
\newcommand\trimBottom{1.0}

\begin{figure}[!tb]
  \centering
  
  \subfloat{%
  \includegraphics[clip, trim=0.2cm \trimBottom cm  0.1cm 0cm, width=\widthFig]{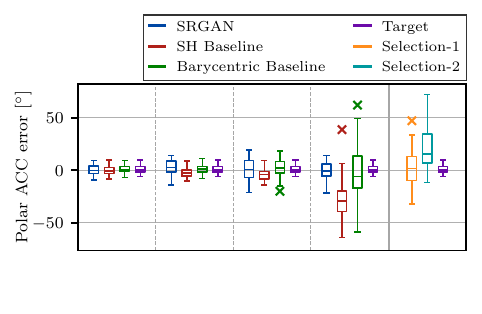}
} \vspace{-0.35cm}
  
  \subfloat{%
  \includegraphics[clip, trim=0.2cm \trimBottom cm 0.1cm \trimTop cm, width=\widthFig]{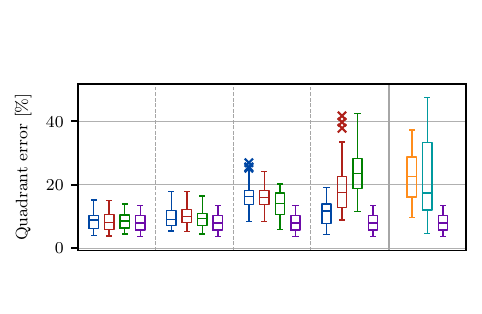}
} \vspace{-0.35cm}
  
  \subfloat{%
  \includegraphics[clip, trim=0.2cm 0cm 0.1cm \trimTop cm, width=\widthFig]{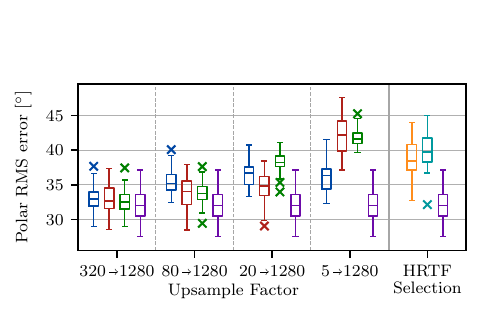}
} 
\vspace{-0.2cm}
 \caption{Results from the model-based perceptual evaluation.}
\label{fig:loc_eval}
\end{figure}

\subsection{LSD metric evaluation}
\label{sec:SD_metric_evaluation}

The \ac{LSD} metric, defined in \cref{eq:SD}, can be calculated for every measurement source position and then averaged over all the source positions. \cref{tab:sd_error} and \cref{fig:sd_error} show the average results for this \ac{LSD} evaluation over the 45 subjects in the test set. In \cref{fig:sd_error}, it is clear to see the benefit of using the proposed \ac{SRGAN} over the barycentric interpolation (Baseline-2) when the input \ac{HRTF} is spatially sparse. The \ac{SRGAN} outperforms the barycentric method when the input contains 20 or fewer different \ac{IR} source positions. The most striking result is upsampling 5 positions to 1280 positions (5  $\,\rightarrow$  1280 network), where the barycentric method only achieves an \ac{LSD} error of 7.30. This makes sense as barycentric interpolation can only average the three closest \acp{IR}, enclosing the interpolated point. The further away these closest points are that form the enclosing triangle, the more inaccurate the barycentric approach becomes. 

In terms of the \ac{SH} interpolation baseline (Baseline-1), it performs slightly worse than barycentric interpolation across all upsampling factors and never outperforms the \ac{SRGAN}. This result is likely due to the distribution of the source positions not being uniformly spaced around the sphere after the gnomonic equiangular projection. It is expected that the performance of \ac{SH} and barycentric interpolation will be comparable as both methods effectively rely on a weighted sum of existing points to generate new ones having no prior knowledge of the \ac{HRTF} data.  

The results of \ac{HRTF} selection (Baseline-3) are also shown to give poor performance in terms of \ac{LSD} error when compared against the proposed method and barycentric interpolation (Baseline-2). What is interesting is that the \ac{HRTF} Selection-1 method slightly outperforms barycentric interpolation when only 5 source positions are used as the low-resolution input (with a mean \ac{LSD} error of 6.96 for Selection-1 compared with 7.30 for barycentric), however, as is expected, Selection-2 does perform worse (where the mean \ac{LSD} error is 8.20 for Selection-2 compared with 7.30 for barycentric). 

To better understand where these errors occur for the proposed \ac{SRGAN} and barycentric interpolation, an illustrative example of a random individual in the test set (SubjectID 868) is given in \cref{fig:sd_error_all_nodes}, where the \ac{LSD} errors for all the interpolated source positions are shown for the different levels of upsampling. The \ac{LSD} errors for Baseline-3 are also given in \cref{fig:sd_error_all_nodes_min_max} for the same subject.  

It can be seen in \crefsub{fig:sd_error_all_nodes_320} for upsampling factor 320~$\rightarrow\,$1280 that the source positions with larger errors over 7~dB are the same for both methods. However, as the input becomes more spatially sparse, the errors between both methods start to diverge, which can be seen in \crefsub{fig:sd_error_all_nodes_20} and \crefsub{fig:sd_error_all_nodes_5}. This is due to the fact that barycentric interpolation is unable to interpolate source positions that are not close to measured points. On the other hand, the proposed \ac{SRGAN} has learnt general patterns across the training data and the relationships between these and the low-resolution input. It is, therefore, able to perform equally across all source positions regardless of where they are located and their proximity to measurement points. This is most clearly seen when comparing the \ac{SRGAN} approach in \crefsub{fig:sd_error_all_nodes_20} and \crefsub{fig:sd_error_all_nodes_5}. The errors are almost identical across all source positions (see also \cref{tab:sd_error}), although the number of original points has been reduced by a factor of four. In contrast, when comparing the barycentric interpolation for the same upsampling levels, the errors increase substantially from \crefsub{fig:sd_error_all_nodes_20} to \crefsub{fig:sd_error_all_nodes_5} at the points where the input measurements have been discarded (e.g. above and below the equator).

\subsection{Model-based perceptual evaluation}

In this section, we use a Bayesian model, Barumerli2022, introduced in \cite{Barumerli2023}, to compare the localisation performance. Unlike \cref{sec:SD_metric_evaluation}, where the \ac{LSD} metric is used to compare the performance of the different methods, the evaluation in this section is able to differentiate the different techniques based on errors that matter to human perception. This is important as some minor errors in the \ac{LSD} could significantly impact human localisation performance. Likewise, some significant errors in the \ac{LSD} may not affect localisation performance nearly as much. The Barumerli2022 model was chosen as it has already been successfully employed in the past for evaluating different binaural rendering methods \cite{Engel2022}.

The Barumerli2022 model was fed features that were obtained from the \acp{DTF}, which is the directional component directly extracted from the HRTF \cite{Middlebrooks1999}. The model's predefined parameters were set to: 
`estimator': \ac{MAP},
`num\_exp': 300,
`sigma\_itd':  0.569,
`sigma\_ild': 1,
`sigma\_spectral': 4,
`sigma\_prior': 11.5,
`sigma\_motor': 10.
The `targ\_az' and `targ\_el' are set to the source positions measured in the \ac{HRTF}.

To perform an effective comparison, the results for the original high-resolution measured \acp{HRTF} are provided as the `Target' results. These `Target' results are the best performance that can be achieved as they effectively compare the localisation performance of the original high-resolution \ac{HRTF} with itself. Therefore, the proposed method and the two other baselines must be benchmarked against the `Target' performance.

The results obtained from the Barumerli2022 model are shown in \cref{tab:perceptual_evaluation} with a graphical representation given in \cref{fig:loc_eval}. \cref{tab:p_acc_error}, \cref{tab:quadrant_error} and \cref{tab:p_rms_error} show the mean and \ac{SD} of the polar accuracy error, quadrant error and \ac{RMS} error, respectively. These metrics \cite{Middlebrooks1999a, Baumgartner2015} can be defined mathematically for $N$ localisation trials, where each target source direction $\phi_i$ has an associated response direction $\Tilde{\phi}_i$, for $i = 1, 2, \dots, N$. If a set of local responses is then defined as $\mathcal{A} = \{i:\text{wrap}|\Tilde{\phi}_i-\phi_i| < 90^{\circ}\}$ the three error metrics are defined as follows: 
\begin{align} 
  \text{Polar Accuracy} &= \frac{\Tilde{\phi}_i-\phi_i}{|\mathcal{A}|} \;,
  \end{align}
 \begin{align} 
   \text{Polar RMS Error} &= \sqrt{\frac{\sum_{i\in\mathcal{A}}(\text{wrap}(\Tilde{\phi}_i-\phi_i))^2 }{|\mathcal{A}|}}\;, 
\\
  \text{Quadrant Error} &= \Big(1 - \frac{|\mathcal{A}|}{N}\Big) \times 100 \;,
\end{align}
where the polar accuracy shows the local response bias in the polar angles for responses within 90$^{\circ}$ of the target. The polar \ac{RMS} error is the aggregated error in the polar dimension for responses within 90$^{\circ}$ of the target. The quadrant error rate corresponds to the percentage of polar errors larger than 90$^{\circ}$ and accounts for top-down and front-back confusions. To avoid highly distorted polar errors on the far left and right sides of the listener, polar and quadrant errors are only defined for targets within a lateral angle $|\theta|~\leq~30^{\circ}$.

It can be seen that when the input provides 320 or 80 source positions, the \ac{SRGAN} performance is comparable to \ac{SH} and barycentric interpretation, only performing marginally worse. However, when the low-resolution \ac{HRTF} contains 20 or fewer source positions, the proposed \ac{SRGAN} performs significantly better than all baselines.

It is also interesting that the \ac{SRGAN} at upsample factor 5~$\,\rightarrow$~1280 outperforms the \ac{SRGAN} at 20~$\,\rightarrow$~1280. This is because even though the \ac{LSD} is worse when only 5 points are used, that does not mean that those spectral differences are perceptually relevant. This shows the need to incorporate these perceptual models into the loss function going forward by overcoming the need for them to be differentiable. It should also be noted that when only 5 source positions are available, the centre position of each panel is selected, and therefore, the source positions are not an exact subset of the 20. This highlights, understandably, the likelihood that certain positions could be more meaningful for localisation when compared with other positions.

\subsection{Limitations and future work}

While the results presented in the previous sections are relatively positive and confirm the potential suitability of the proposed approach for spatially upsampling very sparse \ac{HRTF} measurements, there are some evident limitations which should be addressed in future research.

Firstly, this study didn't assess the impact of the pre-processing transformations on numerical and model-based perceptual metrics. The limitation of the current method is that to achieve optimal performance and reduce the Barycentric interpolation errors while upsampling an existing dataset; the approach currently relies on the model being retrained with the input positions that are closest to any given point on the high-resolution cartesian grid. In future work, it would be good to explore the possibility of passing the position of each point as an additional input feature to the model so that this retaining is not needed.
% Such investigation would allow, for example, to assess the importance of using (rather than discarding) projected points at an elevation angle lower than the measured ones. 

In order to simplify the SRGAN architecture, for this first study, we have disregarded the phase information and used only the magnitude component of each \ac{HRTF}. Upsampled \acp{HRTF} were then reconstructed using minimum-phase approximation and an ITD model. Future research should explore the possibility of including phase information when training the network, and further evaluations should outline how this may result in perceptually relevant improvements.

Another simplification was performed when looking at the number of panels on which the projection was performed during the \ac{HRTF} pre-processing stages. The choice of 5 panels was mainly dictated by the dataset used for the training. Still, it is, of course, possible to extend this to 6 panels, therefore, to include all \acp{HRTF} measured below the subject both as training and output data. Future research could also look at removing this limitation, allowing an arbitrary number (and position) of \acp{HRTF} to be used for the upsampling (currently, this is limited to a minimum of 5 positions, one per panel) and exploring the impact of having non-uniform measurements as a starting point for the upsampling process. 

The design of the loss function could also benefit from further investigations. For example, looking at the weight of each of the two components, ILD and LSD, increasing it for the first in lateral positions and for the second in positions on the sagittal planes. Furthermore, perceptual models could also be employed, but relevant challenges should be addressed first, especially related to the computational complexity of such models.

 It is important to underline that this paper describes the second milestone (after \cite{Siripornpitak2022}) in the process of designing, validating, improving and ultimately releasing this method as an openly available tool. Research following these future directions is already being conducted.
\section{Conclusion} % and Discussion}
\label{sec:conclusion} In this paper, it has been shown that an \ac{SRGAN} can be used effectively for the task of upsampling low-resolution \acp{HRTF}. Furthermore, this work has extended the pilot study from \cite{Siripornpitak2022}, modifying the \ac{SRGAN} so that it can upsample the \acp{HRTF} in \ac{3D} across the entire sphere. It has been demonstrated that when the low-resolution \ac{HRTF} input is very sparse and consists of less than 20 source positions, the \ac{SRGAN} outperforms both \ac{SH} and barycentric interpolation in terms of \ac{LSD} error. The same applies to localisation performance using perceptual models when the low-resolution \ac{HRTF} contains 5 source positions. Therefore, in the case where the low-resolution \ac{HRTF} contains 320 or more source positions, it is preferable to use barycentric interpolation; however, if the \acp{HRTF} are very sparse and contain less than 20 measurements, then the \ac{SRGAN} approach produces significantly lower errors. \ac{SH} interpolation, on the other hand, performs slightly worse than both barycentric interpolation and the \ac{SRGAN}, which is likely due to the distribution of the source positions not being uniformly spaced around the sphere. Non-individual \ac{HRTF} selection also never performs best; however, it does outperform \ac{SH} and barycentric interpolation for very sparse \acp{HRTF}, according to the employed metrics.

In order to reinforce the idea of reproducible research and promote future development and innovation in this specific domain, the complete SRGAN architecture and pre-processing code can be found in our public repository \cite{Hogg2023}.

\ifCLASSOPTIONcaptionsoff
  \newpage
\fi

% \newpage
% \clearpage
% \clearpage
% \pagebreak
% \bibliographystyle{IEEEbib}
% {\setstretch{.95}
\bibliographystyle{IEEEtran}
\bibliography{references/axdstrings,references/AXD}

\begin{thebibliography}{10}
\providecommand{\url}[1]{#1}
\def\UrlFont{\rmfamily}
\providecommand{\newblock}{\relax}
\providecommand{\bibinfo}[2]{#2}
\providecommand\BIBentrySTDinterwordspacing{\spaceskip=0pt\relax}
\providecommand\BIBentryALTinterwordstretchfactor{4}
\providecommand\BIBentryALTinterwordspacing{\spaceskip=\fontdimen2\font plus
\BIBentryALTinterwordstretchfactor\fontdimen3\font minus
  \fontdimen4\font\relax}
\providecommand\BIBforeignlanguage[2]{{%
\expandafter\ifx\csname l@#1\endcsname\relax
\typeout{** WARNING: IEEEtran.bst: No hyphenation pattern has been}%
\typeout{** loaded for the language `#1'. Using the pattern for}%
\typeout{** the default language instead.}%
\else
\language=\csname l@#1\endcsname
\fi
#2}}

\bibitem{Lokki2004}
T.~Lokki, H.~Nironen, S.~Vesa, L.~Savioja, A.~H{\"a}rm{\"a}, and
  M.~Karjalainen, ``Application scenarios of wearable and mobile augmented
  reality audio,'' in \emph{Proc. Audio Eng. Soc. ({AES}) Conv.}, May 2004.

\bibitem{Vickers2021}
D.~Vickers, M.~{Salorio-Corbetto}, S.~Driver, C.~Rocca, Y.~Levtov, K.~Sum,
  B.~Parmar, G.~Dritsakis, J.~Albanell~Flores, D.~Jiang, \emph{et~al.},
  ``Involving children and teenagers with bilateral cochlear implants in the
  design of the {{BEARS}} ({{Both EARS}}) virtual reality training suite
  improves personalization,'' \emph{Front. Digit. Health}, vol.~3, Nov. 2021.

\bibitem{Wightman1989}
F.~L. Wightman and D.~J. Kistler, ``Headphone simulation of free-field
  listening. {{I}}: Stimulus synthesis,'' \emph{J. Acoust. Soc. Am.}, vol.~85,
  no.~2, Feb. 1989.

\bibitem{Blauert1983}
J.~Blauert, \emph{{Spatial hearing: the psychophysics of human sound
  localization}}.\hskip 1em plus 0.5em minus 0.4em\relax {Cambridge, Mass}:
  {MIT Press}, 1983.

\bibitem{Stitt2019}
P.~Stitt, L.~Picinali, and B.~F.~G. Katz, ``Auditory accommodation to poorly
  matched non-individual spectral localization cues through active learning,''
  \emph{Scientific Reports}, vol.~9, no.~1, pp. 1063:1--14, Jan. 2019.

\bibitem{Wenzel1993}
E.~M. Wenzel, M.~Arruda, D.~J. Kistler, and F.~L. Wightman, ``Localization
  using nonindividualized head-related transfer functions,'' \emph{J. Acoust.
  Soc. Am.}, vol.~94, no.~1, pp. 111--123, July 1993.

\bibitem{Moller1996}
H.~M{\o}ller, M.~F. S{\o}rensen, C.~B. Jensen, and D.~Hammersh{\o}i, ``Binaural
  technique: Do we need individual recordings?'' \emph{J. Audio Eng. Soc.
  ({AES})}, vol.~44, no.~6, pp. 451--469, June 1996.

\bibitem{Kahana2006}
Y.~Kahana and P.~A. Nelson, ``Numerical modelling of the spatial acoustic
  response of the human pinna,'' \emph{J. of Sound and Vibration}, vol. 292,
  no.~1, pp. 148--178, Apr. 2006.

\bibitem{Simon2016}
L.~S.~R. Simon, N.~Zacharov, and B.~F.~G. Katz, ``Perceptual attributes for the
  comparison of head-related transfer functions,'' \emph{J. Acoust. Soc. Am.},
  vol. 140, no.~5, pp. 3623--3632, Nov. 2016.

\bibitem{Werner2016}
S.~Werner, F.~Klein, T.~Mayenfels, and K.~Brandenburg, ``A summary on acoustic
  room divergence and its effect on externalization of auditory events,'' in
  \emph{Proc. Int. Conf. on Quality of Multimedia Experience ({QoMEX})}, June
  2016, pp. 1--6.

\bibitem{Engel2019}
I.~Engel, D.~L. Alon, P.~W. Robinson, and R.~Mehra, ``The effect of generic
  headphone compensation on binaural renderings,'' in \emph{Proc. Audio Eng.
  Soc. ({AES}) Int. Conf. on Immersive and Interactive Audio}, Mar. 2019.

\bibitem{Cuevas-Rodriguez2021}
M.~{Cuevas-Rodriguez}, D.~{Gonzalez-Toledo}, A.~{Reyes-Lecuona}, and
  L.~Picinali, ``Impact of non-individualised head related transfer functions
  on speech-in-noise performances within a synthesised virtual environment,''
  \emph{J. Acoust. Soc. Am.}, vol. 149, no.~4, pp. 2573--2586, Apr. 2021.

\bibitem{Engel2023}
I.~Engel, R.~Daugintis, T.~Vicente, A.~O.~T. Hogg, J.~Pauwels, A.~J. Tournier,
  and L.~Picinali, ``The {{SONICOM HRTF}} dataset,'' \emph{J. Audio Eng. Soc.
  ({AES})}, June 2023.

\bibitem{Ziegelwanger2015a}
H.~Ziegelwanger, W.~Kreuzer, and P.~Majdak, ``{{Mesh2HRTF}}: {{An}} open-source
  software package for the numerical calculation of head-related transfer
  functions,'' in \emph{Proc. Int. Cong. on Sound and Vibration ({ICSV})}, July
  2015, pp. 1--8.

\bibitem{Katz2001}
B.~F.~G. Katz, ``Boundary element method calculation of individual head-related
  transfer function. {{I}}. {{Rigid}} model calculation,'' \emph{J. Acoust.
  Soc. Am.}, vol. 110, no.~5, pp. 2440--2448, Nov. 2001.

\bibitem{Stitt2021}
P.~Stitt and B.~F. Katz, ``Sensitivity analysis of pinna morphology on
  head-related transfer functions simulated via a parametric pinna model,''
  \emph{J. Acoust. Soc. Am.}, vol. 149, no.~4, pp. 2559--2572, Apr. 2021.

\bibitem{Geronazzo2019}
M.~Geronazzo, E.~Peruch, F.~Prandoni, and F.~Avanzini, ``Applying a
  single-notch metric to image-guided head-related transfer function selection
  for improved vertical localization,'' \emph{J. Audio Eng. Soc. ({AES})},
  vol.~67, no.~6, pp. 414--428, June 2019.

\bibitem{Zotkin2003}
{\relax DYN}.~Zotkin, J.~Hwang, R.~Duraiswaini, and L.~S. Davis, ``{{HRTF}}
  personalization using anthropometric measurements,'' in \emph{Proc. {IEEE}
  Workshop on Appl. of Signal Process. to Audio and Acoust. ({WASPAA})}.\hskip
  1em plus 0.5em minus 0.4em\relax {IEEE}, Oct. 2003, pp. 157--160.

\bibitem{Katz2012}
B.~F. Katz and G.~Parseihian, ``Perceptually based head-related transfer
  function database optimization,'' \emph{J. Acoust. Soc. Am.}, vol. 131,
  no.~2, pp. EL99--EL105, Feb. 2012.

\bibitem{Kim2020}
C.~Kim, V.~Lim, and L.~Picinali, ``Investigation into consistency of subjective
  and objective perceptual selection of non-individual head-related transfer
  functions,'' \emph{J. Audio Eng. Soc. ({AES})}, vol.~68, no.~11, pp.
  819--831, Dec. 2020.

\bibitem{Zagala2020}
F.~Zagala, M.~Noisternig, and B.~F. Katz, ``Comparison of direct and indirect
  perceptual head-related transfer function selection methods,'' \emph{J.
  Acoust. Soc. Am.}, vol. 147, no.~5, pp. 3376--3389, May 2020.

\bibitem{Picinali2022}
L.~Picinali and B.~F.~G. Katz, ``System-to-user and user-to-system adaptations
  in binaural audio,'' in {{Sonic}} interactions in virtual environments,'' in
  \emph{Sonic Interactions in Virtual Environments}, M.~Geronazzo and
  S.~Serafin, Eds.\hskip 1em plus 0.5em minus 0.4em\relax Springer, Oct. 2022,
  pp. 121--144.

\bibitem{Carpentier2014}
T.~Carpentier, H.~Bahu, M.~Noisternig, and O.~Warusfel, ``Measurement of a
  head-related transfer function database with high spatial resolution,'' in
  \emph{Proc. {EAA} Forum Acusticum, Eur. Congress on Acoust.}, {Krakow,
  Poland}, Sept. 2014.

\bibitem{Farina2000}
A.~Farina, ``Simultaneous measurement of impulse response and distortion with a
  swept-sine technique,'' in \emph{Proc. Audio Eng. Soc. ({AES}) Conv.}, Feb.
  2000, pp. 1--23.

\bibitem{Zotkin2006}
D.~N. Zotkin, R.~Duraiswami, E.~Grassi, and N.~A. Gumerov, ``Fast head-related
  transfer function measurement via reciprocity,'' \emph{J. Acoust. Soc. Am.},
  vol. 120, no.~4, pp. 2202--2215, Oct. 2006.

\bibitem{Richter2016}
J.-G. Richter, G.~Behler, and J.~Fels, ``Evaluation of a fast {{HRTF}}
  measurement system,'' in \emph{Proc. Audio Eng. Soc. ({AES}) Conv.}, vol.
  140, May 2016, p. 9498.

\bibitem{Zhong2014}
X.-L. Zhong and B.-S. Xie, \emph{Head-Related Transfer Functions and Virtual
  Auditory Display}.\hskip 1em plus 0.5em minus 0.4em\relax {Plantation, FL,
  USA}: InTech, Mar. 2014.

\bibitem{Hartung1999}
K.~Hartung, J.~Braasch, and S.~J. Sterbing, ``Comparison of different methods
  for the interpolation of head-related transfer functions,'' in \emph{Proc.
  Audio Eng. Soc. ({AES}) Conf. on Spatial Sound Reproduction}, Mar. 1999.

\bibitem{Poirier-Quinot2018a}
D.~{Poirier-Quinot} and B.~F.~G. Katz, ``The anaglyph binaural audio engine,''
  in \emph{Proc. Audio Eng. Soc. ({AES}) Conv.}, ser. 144, May 2018.

\bibitem{Cuevas-Rodriguez2019}
M.~{Cuevas-Rodr{\'i}guez}, L.~Picinali, D.~{Gonz{\'a}lez-Toledo}, C.~Garre,
  E.~{de la Rubia-Cuestas}, L.~{Molina-Tanco}, and A.~{Reyes-Lecuona}, ``{{3D}}
  tune-in toolkit: {{An}} open-source library for real-time binaural
  spatialisation,'' \emph{{PLOS} {ONE}}, vol.~14, no.~3, p. e0211899, Mar.
  2019.

\bibitem{Gamper2013}
H.~Gamper, ``Head-related transfer function interpolation in azimuth,
  elevation, and distance,'' \emph{J. Acoust. Soc. Am.}, vol. 134, no.~6, pp.
  EL547--EL553, Dec. 2013.

\bibitem{Evans1998}
M.~J. Evans, J.~A.~S. Angus, and A.~I. Tew, ``Analyzing head-related transfer
  function measurements using surface spherical harmonics,'' \emph{J. Acoust.
  Soc. Am.}, vol. 104, no.~4, pp. 2400--2411, June 1998.

\bibitem{Porschmann2019}
C.~P{\"o}rschmann, J.~M. Arend, and F.~Brinkmann, ``Directional equalization of
  sparse head-related transfer function sets for spatial upsampling,''
  \emph{{IEEE/ACM} Trans. Audio, Speech, Language Process.}, vol.~27, no.~6,
  pp. 1060--1071, June 2019.

\bibitem{Arend2021}
J.~M. Arend, F.~Brinkmann, and C.~P{\"o}rschmann, ``Assessing spherical
  harmonics interpolation of time-aligned head-related transfer functions,''
  \emph{J. Audio Eng. Soc. ({AES})}, vol.~69, no. 1/2, pp. 104--117, Feb. 2021.

\bibitem{Engel2022}
I.~Engel, D.~F.~M. Goodman, and L.~Picinali, ``Assessing {{HRTF}} preprocessing
  methods for {{Ambisonics}} rendering through perceptual models,'' \emph{Acta
  Acust.}, vol.~6, p.~4, Jan. 2022.

\bibitem{Arend2023b}
J.~M. Arend, C.~P{\"o}rschmann, S.~Weinzierl, and F.~Brinkmann,
  ``Magnitude-corrected and time-aligned interpolation of head-related transfer
  functions,'' \emph{{IEEE/ACM} Trans. Audio, Speech, Language Process.},
  vol.~31, pp. 3783--3799, Sept. 2023.

\bibitem{Chen2019}
T.-Y. Chen, T.-H. Kuo, and T.-S. Chi, ``Autoencoding {{HRTFs}} for {{DNN}}
  based {{HRTF}} personalization using anthropometric features,'' in
  \emph{Proc. {IEEE} Int. Conf. on Acoust., Speech and Signal Process.
  ({ICASSP})}, May 2019, pp. 271--275.

\bibitem{Yao2022}
D.~Yao, J.~Zhao, L.~Cheng, J.~Li, X.~Li, X.~Guo, and Y.~Yan, ``An
  individualization approach for head-related transfer function in arbitrary
  directions based on deep learning,'' \emph{{JASA} Express lett. ({JASA-EL})},
  vol.~2, no.~6, p. 064401, June 2022.

\bibitem{Ito2022}
Y.~Ito, T.~Nakamura, S.~Koyama, and H.~Saruwatari, ``Head-related transfer
  function interpolation from spatially sparse measurements using autoencoder
  with source position conditioning,'' in \emph{Proc. Int. Workshop on Acoust.
  Signal Enhancement ({IWAENC})}, Sept. 2022, pp. 1--5.

\bibitem{Duraiswami2004}
R.~Duraiswami, D.~N. Zotkin, and N.~A. Gumerov, ``Interpolation and range
  extrapolation of {{HRTFs}},'' in \emph{Proc. {IEEE} Int. Conf. on Acoust.,
  Speech and Signal Process. ({ICASSP})}, vol.~4, May 2004, pp. iv--45.

\bibitem{Kestler2019}
G.~Kestler, S.~Yadegari, and D.~Nahamoo, ``Head related impulse response
  interpolation and extrapolation using deep belief networks,'' in \emph{Proc.
  {IEEE} Int. Conf. on Acoust., Speech and Signal Process. ({ICASSP})}, May
  2019, pp. 266--270.

\bibitem{Jiang2023}
Z.~Jiang, J.~Sang, C.~Zheng, A.~Li, and X.~Li, ``Modeling individual
  head-related transfer functions from sparse measurements using a
  convolutional neural network,'' \emph{J. Acoust. Soc. Am.}, vol. 153, no.~1,
  pp. 248--259, Jan. 2023.

\bibitem{Tsui2020}
B.~Tsui, W.~A.~P. Smith, and G.~Kearney, ``Low-order spherical harmonic
  {{HRTF}} restoration using a neural network approach,'' \emph{Appl. Sci.},
  vol.~10, no.~17, p. 5764, Jan. 2020.

\bibitem{Donahue2018}
C.~Donahue, J.~McAuley, and M.~Puckette, ``Adversarial audio synthesis,'' in
  \emph{Proc. Int. Joint Conf. on Learning Representations ({ICLR})}, Feb.
  2018.

\bibitem{Eskimez2019}
S.~E. Eskimez, K.~Koishida, and Z.~Duan, ``Adversarial training for speech
  super-resolution,'' \emph{{IEEE} J. Sel. Topics Signal Process.}, vol.~13,
  no.~2, pp. 347--358, May 2019.

\bibitem{Dong2016}
C.~Dong, C.~C. Loy, K.~He, and X.~Tang, ``Image super-resolution using deep
  convolutional networks,'' \emph{{IEEE} Trans. Pattern Anal. Mach. Intell.},
  vol.~38, no.~2, pp. 295--307, Feb. 2016.

\bibitem{Ledig2017}
C.~Ledig, L.~Theis, F.~Huszar, J.~Caballero, A.~Cunningham, A.~Acosta,
  A.~Aitken, A.~Tejani, J.~Totz, Z.~Wang, \emph{et~al.}, ``Photo-realistic
  single image super-resolution using a generative adversarial network,'' in
  \emph{Proc. {IEEE} Conf. Comput. Vis. Pattern Recognit. ({CVPR})}, July 2017.

\bibitem{Schawinski2017}
K.~Schawinski, C.~Zhang, H.~Zhang, L.~Fowler, and G.~K. Santhanam, ``Generative
  adversarial networks recover features in astrophysical images of galaxies
  beyond the deconvolution limit,'' \emph{Mon. Not. Royal Astron. Soc. Lett.},
  vol. 467, no.~1, pp. L110--L114, May 2017.

\bibitem{Goodfellow2014}
I.~Goodfellow, J.~{Pouget-Abadie}, M.~Mirza, B.~Xu, D.~{Warde-Farley},
  S.~Ozair, A.~Courville, and Y.~Bengio, ``Generative adversarial nets,'' in
  \emph{Proc. Neural Inform. Process. Conf}, vol.~2.\hskip 1em plus 0.5em minus
  0.4em\relax {Curran Associates, Inc.}, Dec. 2014, pp. 2672--2680.

\bibitem{Siripornpitak2022}
P.~Siripornpitak, I.~Engel, I.~Squires, S.~J. Cooper, and L.~Picinali,
  ``Spatial up-sampling of {{HRTF}} sets using generative adversarial networks:
  {{A}} pilot study,'' \emph{Front. in Signal Process.}, vol.~2, Aug. 2022.

\bibitem{Ronchi1996}
C.~Ronchi, R.~Iacono, and P.~S. Paolucci, ``The ``cubed sphere'': A new method
  for the solution of partial differential equations in spherical geometry,''
  \emph{J. Computational Physics}, vol. 124, no.~1, pp. 93--114, Mar. 1996.

\bibitem{Rancic1996}
M.~Ran{\v c}i{\'c}, R.~J. Purser, and F.~Mesinger, ``A global shallow-water
  model using an expanded spherical cube: {{Gnomonic}} versus conformal
  coordinates,'' \emph{J. of Royal Meteorological Soc.}, vol. 122, no. 532, pp.
  959--982, Apr. 1996.

\bibitem{Purser1998}
R.~J. Purser and M.~Ran{\v c}ci{\'c}, ``Smooth quasi-homogeneous gridding of
  the sphere,'' \emph{J. of Royal Meteorological Soc.}, vol. 124, no. 546, pp.
  637--647, Jan. 1998.

\bibitem{Purser2011}
\BIBentryALTinterwordspacing
R.~Purser and M.~Rancic, ``A standardized procedure for the derivation of
  smooth and partially overset grids on the sphere, associated with polyhedra
  that admit regular griddings of their surfaces. {{Part I}}: {{Mathematical}}
  principles of classification and construction,'' {National Weather Service
  (NWS)}, {{NOAA}}/{{NCEP Office Note}} 460, Dec. 2011. [Online]. Available:
  \url{https://www.emc.ncep.noaa.gov/officenotes/FullTOC.html}
\BIBentrySTDinterwordspacing

\bibitem{Putman2007}
W.~M. Putman and S.-J. Lin, ``Finite-volume transport on various cubed-sphere
  grids,'' \emph{J. Computational Physics}, vol. 227, no.~1, pp. 55--78, Nov.
  2007.

\bibitem{Zhou2020}
J.~Zhou, G.~Cui, S.~Hu, Z.~Zhang, C.~Yang, Z.~Liu, L.~Wang, C.~Li, and M.~Sun,
  ``Graph neural networks: {{A}} review of methods and applications,'' \emph{AI
  Open}, vol.~1, pp. 57--81, Jan. 2020.

\bibitem{Hogg2023b}
A.~Hogg, H.~Liu, M.~Jenkins, and L.~Picinali, ``Exploring the impact of
  transfer learning on {{GAN-based HRTF}} upsampling,'' in \emph{Proc. {EAA}
  Forum Acusticum, Eur. Congress on Acoust.}, Sept. 2023.

\bibitem{Andreopoulou2022}
A.~Andreopoulou and B.~F.~G. Katz, ``Perceptual impact on localization quality
  evaluations of common pre-processing for non-individual head-related transfer
  functions,'' \emph{J. Audio Eng. Soc. ({AES})}, vol.~70, no.~5, pp. 340--354,
  May 2022.

\bibitem{Jung2019}
J.-H. Jung, C.~S. Konor, and D.~Randall, ``Implementation of the vector
  vorticity dynamical core on cubed sphere for use in the quasi-3-{{D}}
  multiscale modeling framework,'' \emph{J. Adv. Model. Earth Syst.}, vol.~11,
  no.~3, pp. 560--577, Feb. 2019.

\bibitem{Kalman1960}
R.~E. Kalman, ``A new approach to linear filtering and prediction problems,''
  \emph{Trans. of the ASME J. of Basic Engineering}, vol.~82, no. Series D, pp.
  35--45, Mar. 1960.

\bibitem{hogg2019}
A.~O.~T. Hogg, P.~A. Naylor, and {\relax Christine}.~Evers, ``Speaker change
  detection using fundamental frequency with application to multi-talker
  segmentation,'' in \emph{Proc. {IEEE} Int. Conf. on Acoust., Speech and
  Signal Process. ({ICASSP})}, May 2019.

\bibitem{Hogg2021d}
A.~O.~T. Hogg, C.~Evers, A.~H. Moore, and P.~A. Naylor, ``Overlapping speaker
  segmentation using multiple hypothesis tracking of fundamental frequency,''
  \emph{{IEEE/ACM} Trans. Audio, Speech, Language Process.}, vol.~29, pp.
  1479--1490, Mar. 2021.

\bibitem{Zwillinger2018}
D.~Zwillinger, Ed., \emph{{{CRC}} Standard Mathematical Tables and Formulas},
  33rd~ed.\hskip 1em plus 0.5em minus 0.4em\relax {Boca Raton}: {Chapman and
  Hall/CRC}, Jan. 2018.

\bibitem{Weyn2020}
J.~A. Weyn, D.~R. Durran, and R.~Caruana, ``Improving {{Data-Driven Global
  Weather Prediction Using Deep Convolutional Neural Networks}} on a {{Cubed
  Sphere}},'' \emph{J. Adv. Model. Earth Syst.}, vol.~12, no.~9, p. e02109,
  Aug. 2020.

\bibitem{He2015a}
K.~He, X.~Zhang, S.~Ren, and J.~Sun, ``Delving deep into rectifiers: Surpassing
  human-level performance on {{ImageNet}} classification,'' in \emph{Proc.
  {IEEE} Int. Conf. on Comput. Vision ({ICCV})}, Dec. 2015, pp. 1026--1034.

\bibitem{Zheng2015}
H.~Zheng, Z.~Yang, W.~Liu, J.~Liang, and Y.~Li, ``Improving deep neural
  networks using softplus units,'' in \emph{Proc. Int. Joint Conf. on Neural
  Networks ({IJCNN})}, July 2015, pp. 1--4.

\bibitem{Dubey2022a}
S.~R. Dubey, S.~K. Singh, and B.~B. Chaudhuri, ``Activation functions in deep
  learning: {{A}} comprehensive survey and benchmark,'' \emph{Neurocomputing},
  vol. 503, pp. 92--108, Sept. 2022.

\bibitem{Radford2015}
A.~Radford, L.~Metz, and S.~Chintala, ``Unsupervised representation learning
  with deep convolutional generative adversarial networks,'' in \emph{Proc.
  Int. Joint Conf. on Learning Representations ({ICLR})}, Y.~Bengio and
  Y.~LeCun, Eds., Nov. 2015.

\bibitem{Gutierrez-Parera2022}
P.~{Gutierrez-Parera}, J.~J. Lopez, J.~M. {Mora-Merchan}, and D.~F. Larios,
  ``Interaural time difference individualization in {{HRTF}} by scaling through
  anthropometric parameters,'' \emph{{EURASIP} J. on Audio, Speech, and Music
  Process.}, vol. 2022, no.~1, p.~9, May 2022.

\bibitem{McKenzie2019}
T.~McKenzie, D.~T. Murphy, and G.~Kearney, ``Interaural level difference
  optimization of binaural ambisonic rendering,'' \emph{Appl. Sci.}, vol.~9,
  no.~6, p. 1226, Jan. 2019.

\bibitem{Mehrgardt1977}
S.~Mehrgardt and V.~Mellert, ``Transformation characteristics of the external
  human ear,'' \emph{J. Acoust. Soc. Am.}, vol.~61, no.~6, pp. 1567--1576, June
  1977.

\bibitem{Majdak2022}
\BIBentryALTinterwordspacing
P.~Majdak, ``{{ARI HRTF}} database,'' June 2022. [Online]. Available:
  \url{http://www.kfs.oeaw.ac.at/hrtf}
\BIBentrySTDinterwordspacing

\bibitem{Pauwels2023}
J.~Pauwels, ``The {{Hartufo}} toolkit for machine learning with {{HRTF}}
  data,'' in \emph{Proc. Audio Eng. Soc. ({AES}) Conf. on Spatial and Immersive
  Audio}, Aug. 2023.

\bibitem{AES2022}
``{{AES69-2015}}: {{AES}} standard for file exchange - {{Spatial}} acoustic
  data file format,'' Sept. 2022.

\bibitem{Liaw2018}
R.~Liaw, E.~Liang, R.~Nishihara, P.~Moritz, J.~E. Gonzalez, and I.~Stoica,
  ``Tune: A research platform for distributed model selection and training,''
  \emph{arXiv preprint arXiv:1807.05118}, July 2018.

\bibitem{Kingma2015}
D.~P. Kingma and L.~J. Ba, ``Adam: {{A Method}} for {{Stochastic
  Optimization}},'' in \emph{Proc. Int. Joint Conf. on Learning Representations
  ({ICLR})}, May 2015.

\bibitem{Gulrajani2017}
I.~Gulrajani, F.~Ahmed, M.~Arjovsky, V.~Dumoulin, and A.~Courville, ``Improved
  training of wasserstein {{GANs}},'' in \emph{Proc. Neural Inform. Process.
  Conf}, ser. {{NIPS}}'17.\hskip 1em plus 0.5em minus 0.4em\relax {Red Hook,
  NY, USA}: {Curran Associates Inc.}, Dec. 2017, pp. 5769--5779.

\bibitem{Arend2021a}
J.~M. Arend, F.~Brinkmann, and C.~P{\"o}rschmann, ``Assessing spherical
  harmonics interpolation of time-aligned head-related transfer functions,''
  \emph{J. Audio Eng. Soc. ({AES})}, vol.~69, no. 1/2, pp. 104--117, Feb. 2021.

\bibitem{Hogg2023}
\BIBentryALTinterwordspacing
A.~O.~T. Hogg, J.~Mads, and H.~Liu, 2023. [Online]. Available:
  \url{https://github.com/ahogg/HRTF-upsampling-with-a-generative-adversarial-network-using-a-gnomonic-equiangular-projection}
\BIBentrySTDinterwordspacing

\bibitem{Barumerli2023}
R.~Barumerli, P.~Majdak, M.~Geronazzo, D.~Meijer, F.~Avanzini, and
  R.~Baumgartner, ``A {{Bayesian}} model for human directional localization of
  broadband static sound sources,'' \emph{Acta Acust.}, vol.~7, p.~12, May
  2023.

\bibitem{Middlebrooks1999}
J.~C. Middlebrooks, ``Individual differences in external-ear transfer functions
  reduced by scaling in frequency,'' \emph{J. Acoust. Soc. Am.}, vol. 106,
  no.~3, pp. 1480--1492, Sept. 1999.

\bibitem{Middlebrooks1999a}
------, ``Virtual localization improved by scaling nonindividualized
  external-ear transfer functions in frequency,'' \emph{J. Acoust. Soc. Am.},
  vol. 106, no. 3 Pt 1, pp. 1493--1510, Sept. 1999.

\bibitem{Baumgartner2015}
R.~Baumgartner, P.~Majdak, and B.~Laback, ``Modeling sound-source localization
  in sagittal planes for human listeners,'' \emph{J. Acoust. Soc. Am.}, vol.
  136, no.~2, pp. 791--802, Sept. 2015.

\end{thebibliography}
% }

%  \vfill
% \vspace{-0.70cm}
\begin{IEEEbiography}[{\includegraphics[width=1in,height=1.25in,clip,keepaspectratio]{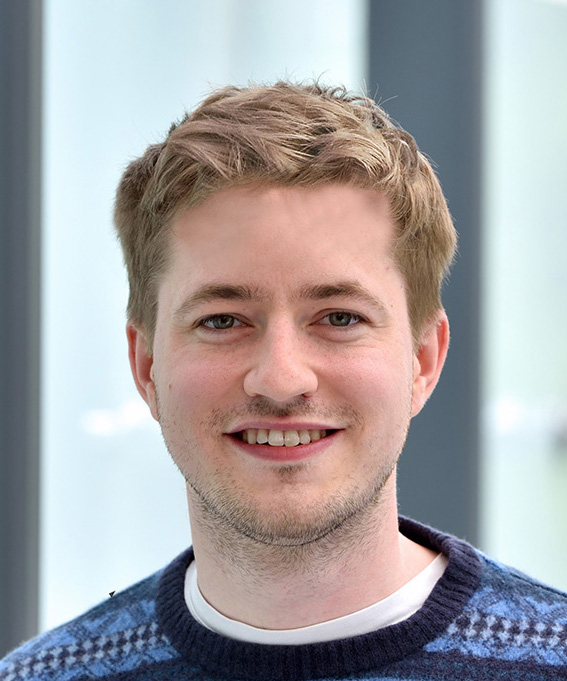}}]{Aidan O. T. Hogg} is a Lecturer in Computer Science at the Centre for Digital Music (C4DM) at Queen Mary University of London. He received an M.Eng. degree in electronic and information engineering and a PhD degree from Imperial College London in 2017 and 2022, respectively. He previously worked as a Research Associate in spatial audio and virtual reality with the Audio Experience Design group at Imperial College London, where he is now an Honorary Research Associate. He has also worked in various engineering roles with Broadcom, Dialog Semiconductor, and Nuance Communications. His current research focuses on using deep learning to capture head-related transfer functions and, more generally, spatial acoustics and immersive audio. Other research interests include speaker diarization and statistical signal processing for audio applications. More information about current research projects can be found here: \url{https://aidanhogg.uk/}
\end{IEEEbiography}
% \vspace{-0.70cm}

\begin{IEEEbiography}[{\includegraphics[width=1in,height=1.25in,clip]{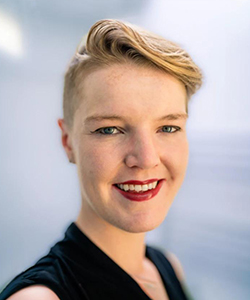}}]{Mads Jenkins} received BSc degrees in Computer Science and in Economics from the Massachusetts Institute of Technology in 2017, and received an MSc degree in Advanced Computing from Imperial College London in 2022. She conducted this spatial audio project with the Audio Experience Design group at Imperial College London, and now works as a machine learning engineer at Cohere, where she develops large language models for conversational applications.
\end{IEEEbiography}
% \vspace{-0.70cm}

\begin{IEEEbiography}[{\includegraphics[width=1in,height=1.25in,clip]{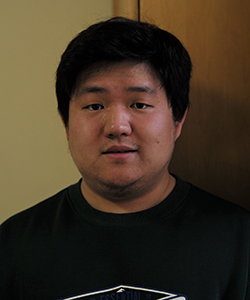}}]{He Liu} received the BA (Mod) Honors degree in Computer Science from Trinity College Dublin in 2021 and MSc degree in Advanced Computing from Imperial College London in 2022. He has worked on projects related to machine learning and deep learning, including human position recognition and video/audio processing.
\end{IEEEbiography}
% \vspace{-0.70cm}

\begin{IEEEbiography}[{\includegraphics[width=1in,height=1.25in,clip]{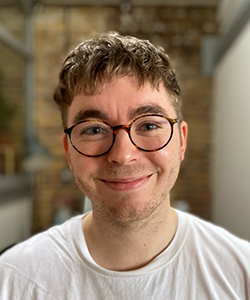}}]{Isaac Squires} is a Ph.D. student in the Dyson School of Design Engineering at Imperial College London. He is supervised by Dr. Samuel J. Cooper, and his research focuses on machine-learning-driven characterization and optimization of battery materials, alongside electrochemical modelling of Li-ion batteries.
\end{IEEEbiography}
% \vspace{-0.70cm}

\begin{IEEEbiography}[{\includegraphics[width=1in,height=1.25in,clip]{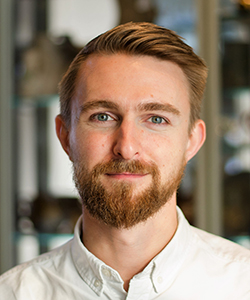}}]{Samuel J. Cooper} is a Senior Lecturer in the Design of Energy Materials at Imperial College London. His team focus on the development of open-source tools for the characterisation and optimisation of energy storage devices using simulations and machine learning. https://tldr-group.github.io/
\end{IEEEbiography}
% \vspace{-0.70cm}

\begin{IEEEbiography}[{\includegraphics[width=1in,height=1.25in,clip,keepaspectratio]{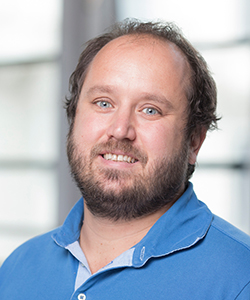}}]{Lorenzo Picinali} is a Reader in Audio Experience Design at Imperial College London. In the past years he worked in Italy, France, and the UK on projects related with 3D binaural sound rendering, interactive applications for visually and hearing impaired individuals, audiology and hearing aids technology, audio and haptic interaction and, more in general, acoustical virtual and augmented reality. More information about the projects in which Lorenzo is involved can be found here \url{https://www.axdesign.co.uk/}
\end{IEEEbiography}

\vfill

\end{document}